\documentclass[lettersize,journal]{IEEEtran}

\usepackage[utf8]{inputenc}
\usepackage{adjustbox}
\usepackage[ruled,vlined]{algorithm2e}
\usepackage{amsfonts}
\usepackage{amsmath}
\usepackage{amssymb, latexsym}
\usepackage{amstext}
\usepackage{amsthm}
\usepackage{array,booktabs,arydshln}
\usepackage{blindtext}
\usepackage{caption} 
\usepackage{subcaption} 
\usepackage{color}
\usepackage{enumitem}
\usepackage{float}
\usepackage{graphicx}
\usepackage{hhline}
\graphicspath{ {./Figures/} }
\usepackage[colorinlistoftodos]{todonotes}
\usepackage{mathtools}
\usepackage{multirow}
\usepackage{nccmath}
\usepackage{textcomp}
\usepackage{url}
\usepackage{verbatim}
\usepackage{xcolor,colortbl}
\usepackage{xfrac} 
\DeclareMathOperator*{\argmin}{argmin} 

\begin{document}
\title{Sonar Point Cloud Processing to Identify Sea Turtles by Pattern Analysis}

\author{Dror~Kipnis,
    Yaniv Levy,
	and~Roee~Diamant,~\IEEEmembership{Senior Member,~IEEE}
	\thanks{
		D.~Kipnis and R.~Diamant are with the Hatter Department of Marine Technologies, University of Haifa, Israel.
		Email: dkipnis@campus.haifa.ac.il, roee.d@univ.haifa.ac.il
		Y. Levy is with the Morris Kahn Marine Research Station, Dep. of Marine Biology, Leon H. Charney School of Marine Sciences, University of Haifa, Haifa 3498838, Israel, and the Israel Sea Turtle Rescue Center, National Nature and Parks Authority, Beit Yannai National Park, Israel.
		
		This work was funded in part by the Israeli Ministry of Energy, Action on Environmental Impact Assessment, under grant 219-17-013.}
}

\maketitle

	\begin{abstract}
		Abundant in coastal areas, sea turtles are affected by high-intensity acoustic anthropogenic sounds. In this paper, we offer a pattern analysis-based detection approach to serve as a warning system for the existence of nearby sea turtles. We focus on the challenge of overcoming the low signal-to-clutter ratio (SCR) caused by reverberations. Assuming that, due to low SCR, target reflections within the point cloud are received in groups, our detector searches for patterns through clustering to identify possible 'blobs' in the point cloud of reflections, and to classify them as either clutter or a target. Our unsupervised clustering is based on geometrical and spectral constraints over the blob's member relations. In turn, the classification of identified blobs as either a target or clutter is based on features extracted from the reflection pattern. To this end, assuming reflections from a sea turtle are stable but include spectral diversity due to distortions within the turtles body, we quantify the stability of the blob's members and their spectral entropy. We test our detector in both modeled simulations, and at sea, for the detection of sea turtles released after rehabilitation. The results show robustness to highly-fluctuating target intensity and ability to detect at low SCR.	
	\end{abstract}



\section{Introduction}\label{sec: Introduction}

Due to the increase in marine anthropogenic activities, it was found that megafauna are threatened by high acoustic intensities with both physical and behavioral impacts \cite{nelms2016seismic}.
While marine mammals and pelagic fish can quickly escape high-intensity acoustic sources, this is not the case for sea turtles.
Hence, there is a need for a detection solution that can serve as a warning system, to alert the presence of sea turtles to stop or mitigate disturbance or detriment during marine operations.
This type of system can be applied over vessels performing seismic surveys that may lead to the interruption of marine turtles' normal behaviors \cite{nelms2016seismic}; within areas of sea water intake to desalination factories that can injure turtles \cite{levy2015small}; within the proximity of areas containing hot water from power plants where turtles aggregate \cite{macdonald2012home}; as well within ports and ship channels and dredging zones where turtles are abundant \cite{bolten1994seasonal}.
Identification of sea turtles' presence in their natural habitat can also be an important tool for population monitoring and conservation.
This paper takes a first step towards such a goal, and offers a pattern recognition approach for detecting sea turtles by active acoustics.

Active acoustic detection of sea turtles was studied for commercial echosounders, where target strength (TS) measurements are conducted at a frequency of 200kHz in a tank and in the open-sea \cite{mahfurdz2018green} \cite{perez2013ts}.
Other related studies \cite{jones2017broadband} use active acoustics to detect fish, based on the resonance of their swimming bladder.
However, when operating at frequencies that differ from the resonance or for species without a swimming bladder, acoustic reflection is expected to be less intense.
Detection methods to identify reflected sonar echoes for very low observable (VLO) targets are studied extensively and include methods to reduce the false alarm rate by utilizing information on reverberation and noise power \cite{abraham2002active},  classification of sonar contacts to target or clutter \cite{berg2018classification} \cite{hjelmervik2017sonar}, tracking based on a motion model \cite{davey2007comparison}, or by probabilistic analysis to identify sequences of reflections based on dynamic programming \cite{diamant2019active}.
However, to the best of our knowledge, no consideration has been given to date, to the time varying characteristics of reflections from targets like sea turtles.
In particular, a main challenge of the active sonar detection of sea turtles is to identify reflections originating from a turtle within a point cloud of clutter reflections.
The problem becomes further compounded when detection is omnidirectional and the SCR is low \cite{de2019automatic}.
On top of this, for a swimming sea turtle that changes its orientation and depth with respect to the receiver, the intensity of the acoustic reflections varies over time.
Thus, rather than a continuous reflection pattern, we expect the reflection pattern from the turtle to form clusters, or 'blobs', within the point cloud of received sonar echoes.

In this work, we introduce a novel pattern recognition-based solution for sea turtle presence detection. Our solution considers the expected time-variation of the reflections from sea turtles to reduce the false alarm rate through point cloud clustering and classification.
We start by accumulating a sequence of reflections from the active sonar signals, arranged as a point cloud such that each point is identified by its ping number and estimated range.
Then, spectral characteristics and geometrical relations between the reflection's temporal and spatial patterns are tested against limitations on the animal's speed and size.
This results in the identification of blobs associated with potential targets.
The identified blobs are then classified as either a target or clutter by examining features derived from the statistics of the cluster's members and from the reflections' spectral diversity.
The latter is based on our key assumption that, due to distortions within the turtle's body, the spectral response of reflections originating from a sea turtle is significantly more diverse than that of the clutter.
Based on this assumption, we use the entropy of the reflections' spectral response as a detection metric.
Furthermore, since we expect clutter reflections to be independent while that of a valid blob to show statistical relations, we validate a blob to be related to a target by the stability of its members.
 
Our contribution is three-fold:
\begin{enumerate}
	\item Design of a graph-based clustering solution that accounts for time variations within a point cloud of sonar's reflections.
	\item Utilization of the statistical dependency between the target's reflections as a measure for pattern recognition.
	\item Design of a detection metric that is based on the expected spectral diversity of acoustic reflections originating from a sea turtle.
\end{enumerate}
We explore our method's performance in both simulations and a designated sea trial.
The simulations are based on a physical model for reflections from a sea turtle and involve real clutter collected during multiple sea trials.
The sea trial involved the detection of two sea turtles we released after rehabilitation.
Results show a favorable trade-off between the false alarm and the detection rates, and resilience to low and time-varying SCR.

The structure of this paper is as follows. 
Section \ref{sec: Related work} surveys the related work in the fields of active sonar detection, classification of sonar reflections, and track-before-detect (TkBD) approaches for VLO targets.
Section \ref{sec: System model} describes the system's model and assumptions, as well as a model for acoustic reflections from sea turtles.
Section \ref{sec: Methodology} details our algorithm, starting from the clustering solution, and followed by the classification of clusters.
Section \ref{sec: Performance analysis} reports the simulation and sea trial results, while conclusions are drawn in Section \ref{sec: Conclusion}.

Throughout the paper, we use the following notations: a small letter indicates a scalar; while vectors and matrices are indicated by bold small and uppercase letters, respectively. The entries of a vector or a matrix are indicated by subscript indices.


\section{Related work} \label{sec: Related work}
The common approach for detecting reflection of active sonar in Gaussian noise is by a matched filter (MF) \cite{robey1992cfar}, which is used widely for active sonar applications \cite{seo2019underwater}\cite{berg2018classification}\cite{hjelmervik2017sonar}.
However, since sound propagation in shallow water is considerably affected by reverberations, the MF's performance deviates from its optimum processing gain \cite{abraham2002active}.
A method to detect sonar echoes in a reverberant environment is proposed in \cite{abraham2002active}, where the Page test is applied on the MF output to identify the start and end of a signal. 
Alternatively, for the specific case of linear frequency modulation (LFM) signals, emitted for object detection, a fractional Fourier transform (FrFT) can be applied to separate signals overlapping in time \cite{cowell2010separation}.
An application of FrFT for sonar signal processing is given in \cite{jacob2009applications} for the parameter estimation of chirp signals.
Detection based on simulations and experimental data is illustrated in \cite{yu2017fractional}, and a better accuracy than that of the MF is observed in terms of the delay estimation.
While the above detection methods aim to identify echoes in low SCR and in conditions of high interference; in practice, due to noise fluctuations, high false alarm rates \cite{hjelmervik2017sonar} are reported in real sea environments.

One way to reduce false detections is by classifying the acoustic reflections to separate between a target and clutter.
In \cite{hjelmervik2017sonar}, a machine-learning approach was demonstrated to reduce the false alarm rate (FAR) of active sonar reflections by using the backscatter intensity as a classification feature.
Similarly, the range-bearing neighborhood of MF sample outputs that passed a detection threshold are used in \cite{seo2019underwater} to extract features for classification by support vector machine (SVM).
In \cite{berg2018classification}, a similar feature set is offered as the input of a neural network to classify the sonar data after MF operation.
Alternatively, \cite{de2019automatic} implemented a convolutional neural network (CNN) to classify a target or clutter by the sonar's spectrogram images.
The drawback is the need for training data, which is often hard to obtain due to the spatial and temporal diversity of the underwater acoustic channel.
Unsupervised anomaly detection \cite{stinco2021automatic} is used to overcome this limitation, making it more practical for real world applications.
However, the classification is limited to individual detected instances, and does not exploit the information of the reflection pattern.

Temporal information of the reflection pattern is obtained when emitting a train of sonar pings whose reflections, after being synchronized, are stacked in a reflection matrix. Analyzing such a matrix can dramatically increase detection performance by using the track-before detect (TkBD) framework.
Common approaches to TkBD include dynamic programming, particle filter (PF), probabilistic multi-hypothesis tracking (PMHT), and probabilistic data association (PDA).
Dynamic programming performs tracking over a discrete grid, assuming the target's dynamics can be modeled as a Markov process with
some limitations on its velocity \cite{barniv1985dynamic}.
Although straightforward, it has the drawback of having a very large computational cost. Application for weak target detection by active sonar can be found in \cite{wei2018novel} on simulated data. 
Viterbi-algorithm (VA)-based TkBD for underwater targets is presented in \cite{diamant2019active}, where detection decision are based on the calculated likelihood of optional target paths.
For lower complexity, PF samples the state-space to estimate the probability density function for target reflections \cite{davey2007comparison}.
\newline
Tracking of an underwater moving source by PF is demonstrated in \cite{jing2016detection}, while \cite{northardt2018track} demonstrated a PF-based TkBD algorithm on a multi-target passive sonar scenario.
In turn, the PMHT approach \cite{streit1994maximum} models a combination of target and noise using the expectation maximization (EM) algorithm, and estimates the target's path by a Kalman smoother.
Extensions include the histogram probabilistic multi-hypothesis tracking (H-PMHT) \cite{streit2002multitarget}\cite{vu2013track}, and the Poisson-H-PMHT \cite{gaetjens2017histogram}.
Finally, the PDA approach considers the overall probability of the set of measurements to infer the existence of a target \cite{jauffret1990track}.
Applications of ML-PDA for real bistatic and multistatic sonar data are demonstrated in \cite{willett2005application} and \cite{blanding2007sequential}, respectively.
A comparison of ML-PDA vs. PMHT on synthetic multistatic active-sonar scenarios is presented in \cite{schoenecker2013ml}.
An underlying assumption for these tracking-based methods is that a target exists, and that its probability of detection is roughly constant within the observation time (i.e., the series of pings).
While such an assumption may be valid for large targets such as submarines, this is not the case for small targets as considered here, i.e., a sea-turtle.
This is because of the high impact of orientation changes, caused by the animal's motion, on the intensity of the acoustic reflections.


\section{System model} \label{sec: System model}
Our setup includes an omnidirectional monostatic sonar, composed of a projector and hydrophone pair. 
A number, $N_\mathrm{p}$, of linear frequency modulation chirp signals are emitted at a constant ping rate interval (PRI), $T_\mathrm{PRI}$, such that reflected echoes allow nearly-uniform temporal sampling of the surrounding.
In this work, we set $N_\mathrm{p} = 20$ for simulations, and $N_\mathrm{p} = 37$ for the experiment.
For each emission, we record the reflected signal, such that the analysis is performed in blocks of time-windows equal to $N_\mathrm{p}\cdot T_\mathrm{PRI}$.
In order to utilize the reflections' expected time-variations, we arrange the $N_\mathrm{p}$ reflections as a set of signals $\{ {\bf y}^{(p)} \}_{p=1}^{N_\mathrm{p}}$, attributed by their pulse index $p$.
Our goal is to find target reflection patterns within these signals.

\subsection{Main assumptions}
Our model assumes that, due to the dynamics of the sea turtle, the intensity of the acoustic reflections form 'blobs' along the target's path at the MF output.
We also assume that, due to the complex structure of the sea turtle's body, the acoustic intensity of these blobs' members is frequency dependent.
Under this model, our detector combines classification with tracking. 

We assume that a valid reflection which originates from a sea turtle follows three conditions. First, the detected path should follow boundaries in terms of the turtle's maximum speed.
Specifically, the range associated with a valid reflection must be distanced up to $v_{\mathrm{max}}\cdot T_{\mathrm{PRI}}$~[m] from its previous or following valid reflection.
Second, a valid reflection must show spectral diversity which differs from that of the clutter.
Last, we expect the samples along a valid target's path at the MF output to be statistically dependent. That is, the joint
probability of the acoustic intensity of a pair of valid reflections should be greater than the multiplication of their separate
probabilities.
In terms of the channel model, we focus on the practical case in which 1) the reflections are received at low SCR, such that energy-based detection is not possible, and that 2) there is non-negligible ambient noise; therefore, statistical analysis over a number of emissions is required.

\subsection{A Model for Acoustic Reflections from Sea Turtles} 
Studies on the distribution of acoustic reflections from sea turtles focus on a commercial ecosounder at a carrier frequency of 200kHz \cite{perez2013ts} \cite{mahfurdz2018green}.
Especially for omni-directional detection, this frequency greatly limits the detection range, as the acoustic attenuation above the
100~kHz becomes a dominant factor in the attenuation loss.
Instead, our transmissions are performed at lower frequencies of 7~kHz to 17~kHz and between 48~kHz to 78~kHz, where the acoustic wavelength is on the same order as the dimensions of the target sea turtle.
For this frequency range, the turtle's reflections are characterized as frequency dependent.
Frequency dependence of acoustic reflections is mostly studied for fish \cite{korneliussen2003synthetic} \cite{stanton2010new} \cite{pailhas2010analysis}.
Here, the acoustic reflections primarily originate from the fish's swim bladder, which, arguably, resemble the reflections from a sea turtle's lungs.
The general model for the spectral diversity of acoustic reflections we refer to is described in \cite{love1971measurements}.
For completeness, we now outline this model in detail below.

The frequency dependence of reflections from a finite-sized object can be divided into three regions, depending on the size of the reflecting object relative to the wavelength.
When the acoustic wavelength $\lambda$ is shorter than the object's size, $l$, the reflection is modeled as Rayleigh scattering, in which the TS rapidly increases with the frequency.
Geometric reflection, which is not frequency dependent, occurs at the optical limit when the wavelength is much smaller than the object's size.
Between these two extremes, there is an interference region in which multiple reflections from the object's boundaries interfere, thus creating a diverse frequency selective pattern.
For a well-defined object there is a clear division between the above regions.
For example, for a sphere with a radius of $a$, the interference region is in the range of $1 \leq \frac{2 \pi a}{\lambda} \leq 10$.
However, for complex shapes like a marine animal this division is not clear, and the animal's internal organs are expected to create a diverse interference pattern.
Empirical results \cite{love1971measurements} identify the interference region for fish as $0.7 \leq \frac{l}{\lambda} \leq 200$, where the lower limit may be even smaller.
For our first transmission range, 7-17~kHz, the size of an object that is within the interference region is between,
\begin{gather}\label{eq: interference region}
	l_{min} f_{min} \geq 0.7c \rightarrow l_{min} \geq \frac{0.7c}{f_{min}} = \frac{0.7 \cdot 1500}{7000} = 0.15 \text{m} \\
	l_{max} f_{max} \leq 200c \rightarrow l_{max} \leq \frac{200c}{f_{max}} = \frac{200 \cdot 1500}{17000} = 17.647 \text{m} \;,
\end{gather}
Similarly, for the frequency range of 48~kHz to 78~kHz, the interference region fits a target size between 2~cm and 3.8~m.
We therefore conclude that the acoustic reflections expected from the sea turtle are within the interference region, and a high diversity in the frequency domain is expected.
In contrast, for clutter composed of smaller elements, acoustic reflection is within the Rayleigh scattering region.
Thus, the clutter's frequency response is not expected to exhibit frequency diversity.
Our approach utilizes this difference in the frequency diversity characteristics of the two signals to separate the turtle's reflections from those of the clutter.

\begin{figure}[h] 
	\centering
	\includegraphics[width=0.48\textwidth]{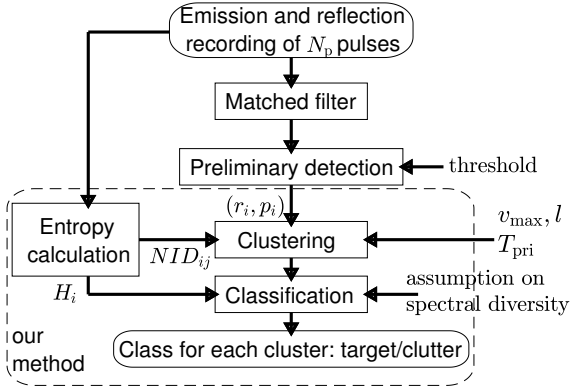}
	\caption{ \label{fig: block diagram1} Block diagram of the processing chain.}
\end{figure}


\section{Methodology}\label{sec: Methodology}

\subsection{Key idea}\label{sec: Key idea}
Our processing chain flow is shown in Fig.~\ref{fig: block diagram1}.
Our solution utilizes our assumption that the acoustic reflections from sea turtles tend to appear as occasional blobs along the turtle's path.
This is due to expected temporal changes in the turtle's orientation and in the volume of its lungs.
The input for our classifier is the set $\{ (r_i, p_i) \}_{i=1}^N$, which is the point cloud after MF thresholding.
Here, $r_i$ represents the range of the point from the sonar, and $p_i$ is the time of acquisition, as illustrated in Fig.~\ref{fig: processing stages}a,b.

Referring to Fig.~\ref{fig: block diagram1}, ours is a two-stage process.
First, we segment target indications into clusters of potential targets.
To this end, we identify reflections that are close in range and time, and thus are more-likely to be from the same object.
As part of the clustering solution, we rank the connectivity between data-points by their mutual information.
Instead of pre-setting the number of clusters, which in our case is not possible, we determine it by minimizing a utility function for the 'cost' of the clustering solution.
Once clusters are formed in the shape of 'blobs', we move to the second phase where we analyze the blob's members to draw features about their statistics, and to determine whether the blob is an indication of a sea turtle-like target.
These features comprise, on the one hand, the spatial, temporal, and statistical connectivity of the acoustic reflections along the blob and, on the other hand, the spectral diversity that is expected for the reflections from a sea turtle.
The former is calculated by summing all the connections within a cluster, while the latter by the entropy of the acoustic reflections at different frequencies.
In doing so, we achieve a processing gain by analyzing groups of reflections.

We note that our solution can manage two cases: one when an exact spectral diversity model is available, and one when no model is available, but spectral diversity is expected.
The first calls for ranking the alignment between the measured diversity and a given template, while the second compares the spectral diversity of the blob to that of the clutter.
In turn, clutter is identified by clusters whose members' connectivity is low, or whose members' entropy is high.
An important notification is that, while we demonstrate our solution for the detection of sea turtles, our solution may also be applied to a variety of targets - ranging from pelagic fish to top predators.
In each case, the user should provide a mobility model (limitation on the speed and movement pattern) and spectral model, if any.

\subsection{Clustering}\label{sec: Clustering} 

The goal of the clustering phase is to segment the thresholded data-points in blobs of possible targets.
These blobs should contain data-points that are close in range and time, and have similar reflection spectrums.
To measure this similarity, we use the normalized information distance (NID) \cite{vinh2010information}, which is calculated for the spectral entropy of the signals associated with reflections from each pair of data-points.
Denote the time domain signal corresponding to the $i$th data-point by ${\bf y}^{(i)}$, and its normalized power spectrum by $\hat {\bf z}^{(i)}_{M \times 1}$.
We calculate the spectral entropy for the $i$th data-point by \cite{misra2004spectral}
\begin{equation}\label{eq: spectral entropy}
	H_i = - \sum_{m=1}^M \hat z^{(i)}_m \mathrm{log_2} \hat z^{(i)}_m,
\end{equation}
where $m$ attributes the frequency bins in which the power spectrum is calculated, and $\hat {\bf z}$ is normalized such that $\bf{\hat z} ^\mathrm T \bf{\hat z} = 1$.
The NID used to quantify spectral similarity between data-points $i$ and $j$ is calculated by
\begin{equation}\label{eq: NID}
	\mathrm{NID}_{ij}( \hat {\bf z}^{(i)}; \hat {\bf z}^{(j)}) = 1 - \frac{ I_{ij}( \hat {\bf z}^{(i)}; \hat {\bf z}^{(j)}) }{ \mathrm{max} [H_i(\hat {\bf z}^{(i)}), H_j(\hat {\bf z}^{(j)})] },
\end{equation}	
where $I_{ij}$ is the mutual information between $\hat {\bf z}^{(i)}$ and $\hat {\bf z}^{(j)}$ obtained by
\begin{equation}\label{eq: mutual information}
	I_{ij}( \hat {\bf z}^{(i)}; \hat {\bf z}^{(j)}) = H_i(\hat {\bf z}^{(i)}) + H_j(\hat {\bf z}^{(j)}) - H_{ij}(\hat {\bf z}^{(i)}, \hat {\bf z}^{(j)}) \;,
\end{equation}
and the term $H_{ij}(\hat {\bf z}^{(i)}, \hat {\bf z}^{(j)})$ is the joint entropy of $\hat {\bf z}^{(i)}$ and $\hat {\bf z}^{(j)}$.
The reason we use the NID measure is because it relates to the mutual information of the reflection spectrum of each pair of data-points \cite{vinh2010information}.
Specifically, a value of NID close to 0 indicates a high level of mutual information, while an NID close to 1 is a case where $\hat {\bf z}^{(i)}, \hat {\bf z}^{(j)}$ are independent.
The range difference of two data-points is denoted by $\Delta r_{ij} = |r_i - r_j|$, and their time-difference by $\Delta t_{ij} = |p_i - p_j |T_\mathrm{PRI}$.
We define the distance of two data-points $i$ and $j$ by
\begin{equation} \label{eq: Delta_ij}
	\Delta_{ij} = \alpha \mathrm{NID}_{ij} + \beta \Delta r_{ij} \mathrm{u}( \Delta r_{ij} - l) +  \tau \Delta t_{ij},
\end{equation}
where $\mathrm{u(.)}$ is the Heaviside step function.
The parameters $\alpha$, $\beta$ and $\tau$ tune the relative importance of the frequency diversity, range difference, and ping difference, respectively.

Our clustering solution is based on the method proposed in \cite{ng2002spectral}, which performs spectral clustering to a pre-defined number of clusters. Here, we perform an additional search to find the most appropriate number of clusters.
This spectral clustering method was selected because it enables the formation of non-convex clusters as expected in our application (i.e., the maneuvering sea turtle does not move in straight lines).
In this method, the data is encoded as a weighted graph, where data-points are the nodes and their edges are weighted according to distances $\Delta_{ij}$ defined in \eqref{eq: Delta_ij}.
We define the weight of the edge between the $i$th and $j$th data-point by
\begin{equation}\label{eq: w_ij}
	w_{ij} = e^{-\Delta_{ij}} \mathrm{u}( v_\mathrm{max} - \frac{2 \Delta r_{ij}}{\Delta t_{ij}} ),
\end{equation}
where the step-function constrains the maximal velocity allowed by the model $v_\mathrm{max}$, and factor 2 is present because the range is two-way.
This processing stage is illustrated in Fig.~\ref{fig: processing stages}c, where the graph edges are marked as dashed lines.
Following \cite{ng2002spectral}, we define a matrix $\bf A$ of entries
\begin{equation} \label{eq: A_ij}
	a_{ij} = 
	\begin{dcases}
		w_{ij}, ~& i \neq j \\
		0, ~& i=j,
	\end{dcases}
\end{equation}
and a diagonal matrix $\bf D$ with entries 
\begin{equation} \label{eq: D_i}
	d_{i} = \sum_{j=1}^N A_{ij}.
\end{equation}
We then calculate the symmetric graph-Laplacian by
\begin{equation} \label{eq: graph-Laplacian}
	{\bf L} = {\bf D}^{-1/2} ({\bf D - A}) {\bf D}^{-1/2}\;.
\end{equation}
The spectral clustering algorithm \cite{ng2002spectral} identifies clusters' centers in the space of the graph-Laplacian's eigenvectors, as detailed in the Appendix.
However, the number of clusters should be provided, as the final stage of \cite{ng2002spectral} relies on k-means clustering.

We estimate the appropriate number of clusters by considering the total 'cost' of the clustering solution.
We note that the term $ {\bf D}^{-1/2} {\bf A} {\bf D}^{-1/2}$ in \eqref{eq: graph-Laplacian} is the ratio of the inner to outer connectivity of each pair of nodes, $0 \leq a_{ij} / \sqrt{d_i d_j} \leq 1$.
Denote the $k$th cluster by an $N \times 1$ binary vector ${\bf b}_k$, whose entries is a binary indicator that is set to 1 if the corresponding data-point belongs to the cluster.
We relate to ${\bf b}_k^\mathrm{T} {\bf L} {\bf b}_k$ as a cost function for forming nodes $i$ and $j$ in the same $k$th cluster.
The appropriate number of clusters, $\hat K$, is determined by minimizing the mean partition cost,
\begin{equation} \label{eq: cost}
	\begin{split}
		& \hat K = \argmin_K \frac{1}{K}\sum_{k=1}^K {\bf b}_k^\mathrm{T} {\bf L} {\bf b}_k + \epsilon K \\
		& \text{s.t. } 	b_i \in [0,1], ~ {\bf b}_k ^\mathrm{T} {\bf b}_l = 0 \quad \forall k \neq l	
	\end{split}	
\end{equation}
where $\epsilon K$ is a penalty for forming 'too many' clusters.
This model-order selection process is illustrated in Fig.~\ref{fig: processing stages}d, while the formed clusters are shown in Fig.~\ref{fig: processing stages}e.

\subsection{Classify clusters}\label{sec: Classify Clusters}
\begin{figure*}[h]
\centering
\subfloat{\includegraphics[width=0.4\textwidth]{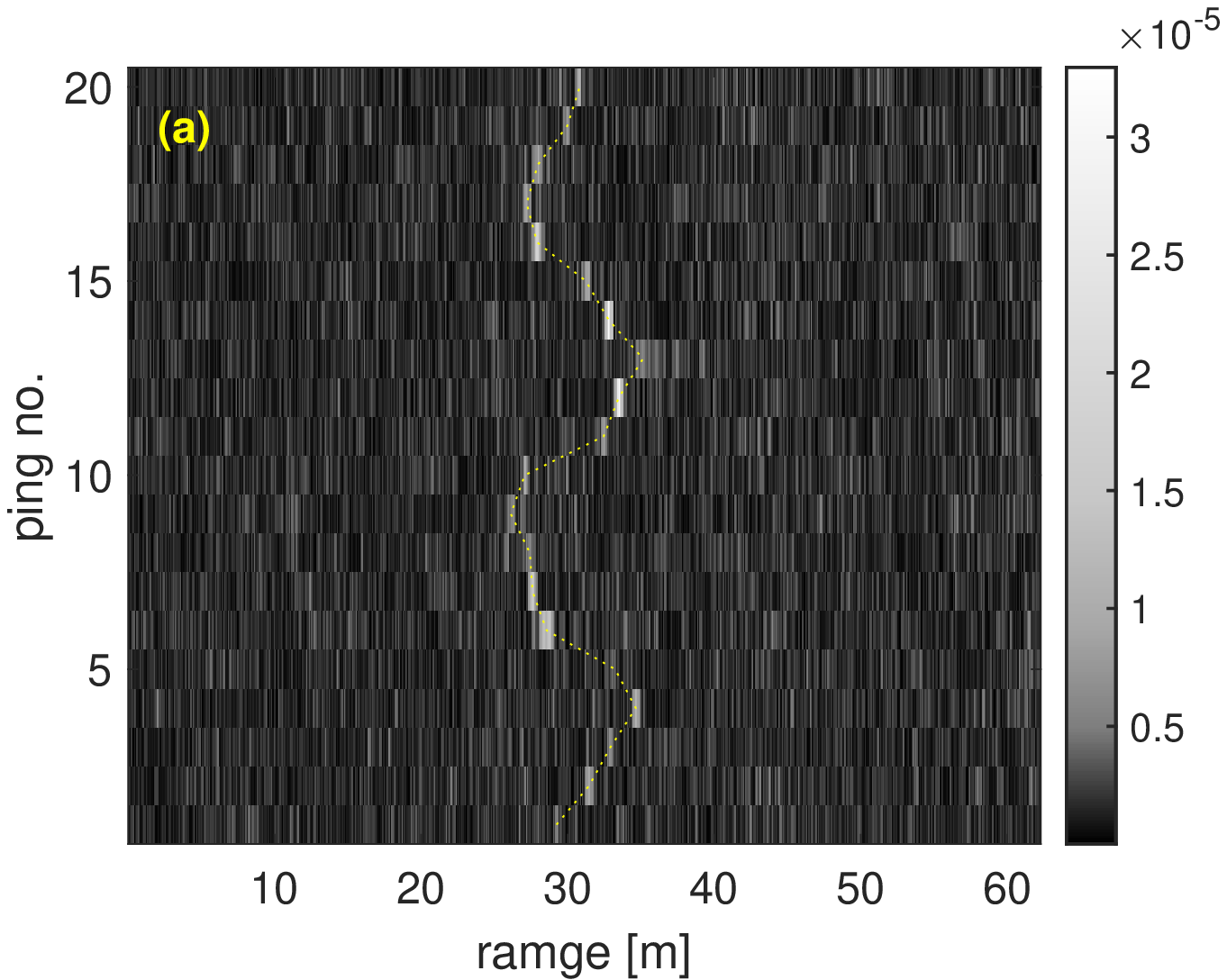}%
\label{fig_a}}
\subfloat{\includegraphics[width=0.4\textwidth]{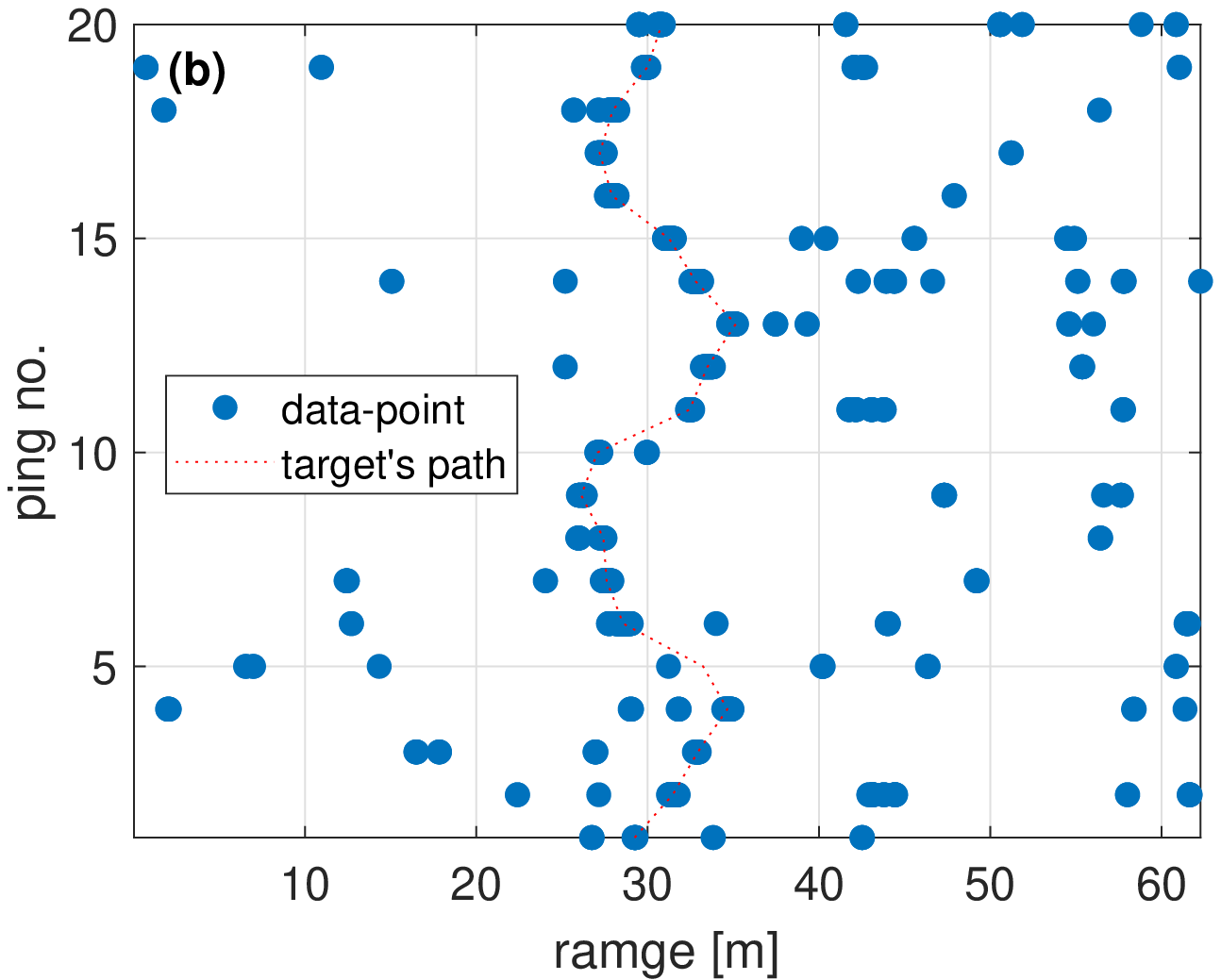}%
\label{fig_b}}
\hfill
\subfloat{\includegraphics[width=0.4\textwidth]{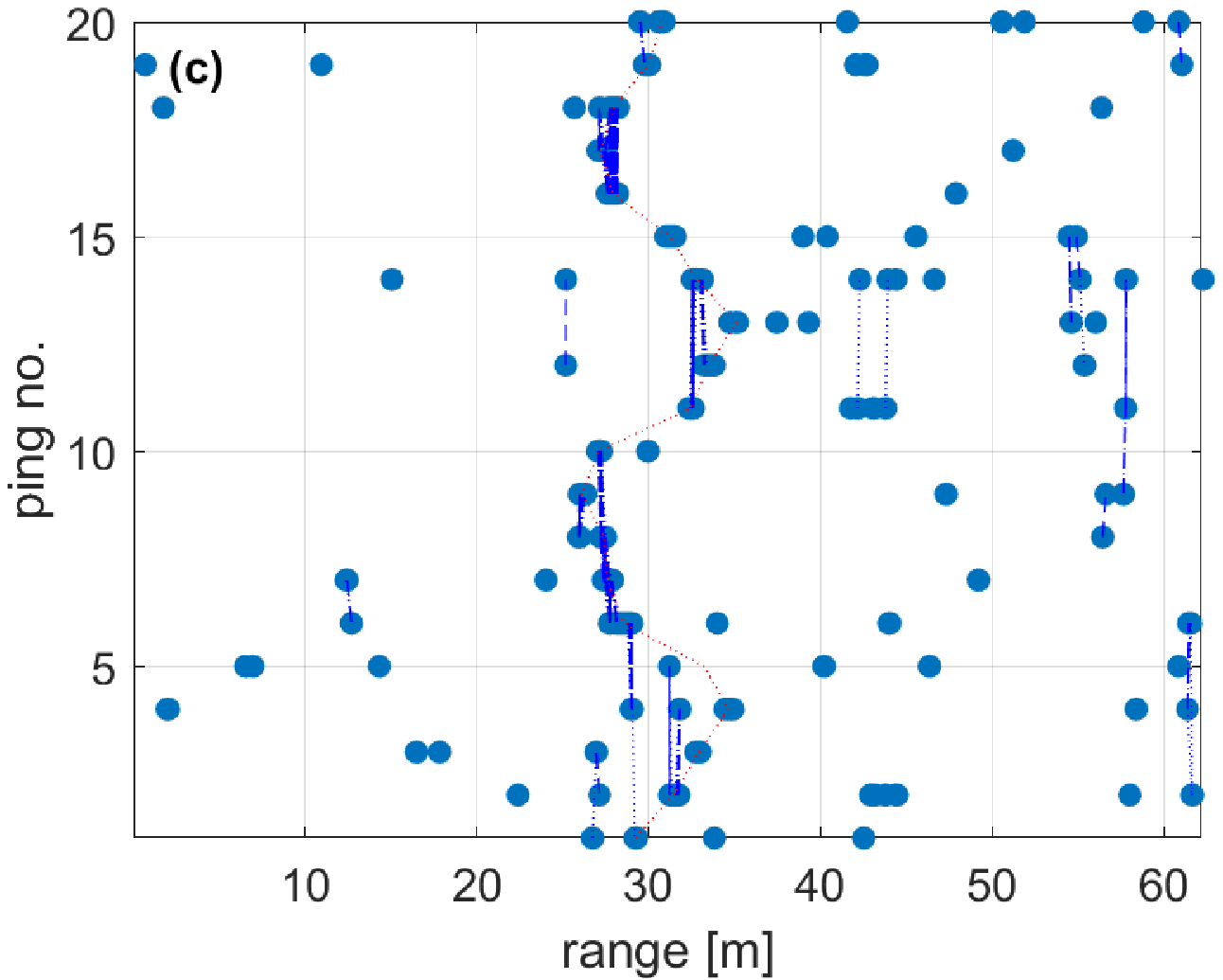}%
\label{fig_c}}
\subfloat{\includegraphics[width=0.4\textwidth]{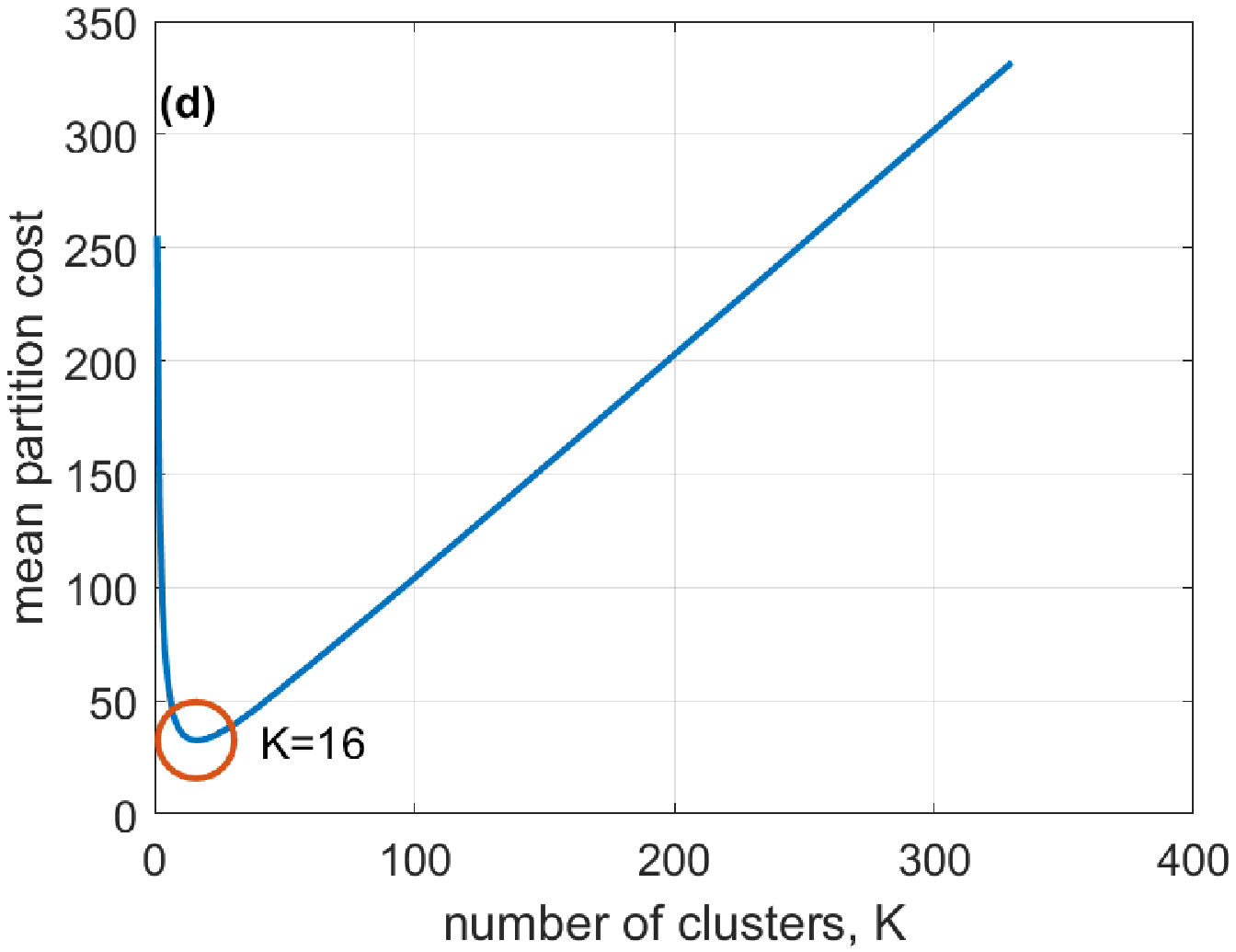}%
\label{fig_d}}
\hfill
\subfloat{\includegraphics[width=0.4\textwidth]{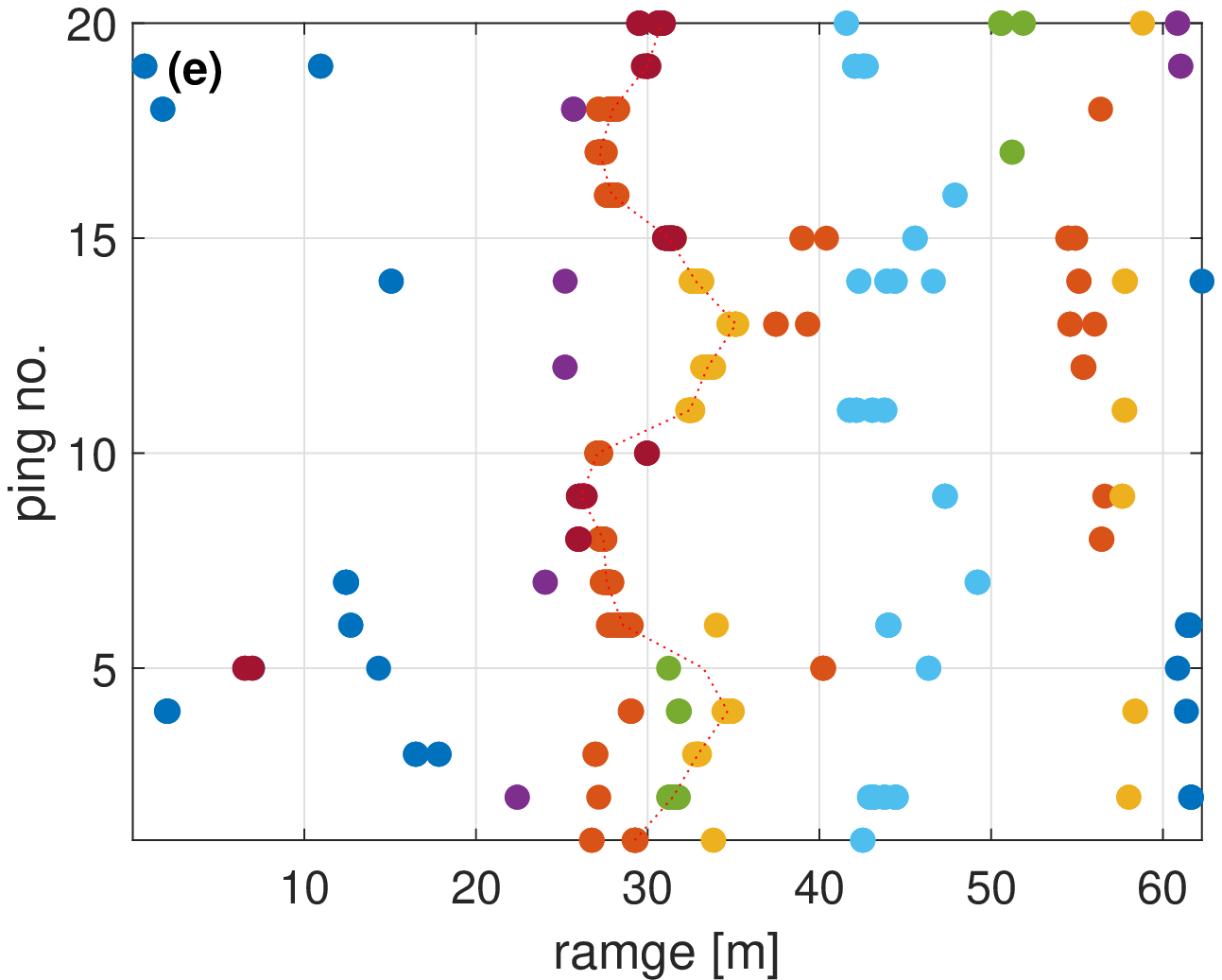}%
\label{fig_e}}
\subfloat{\includegraphics[width=0.4\textwidth]{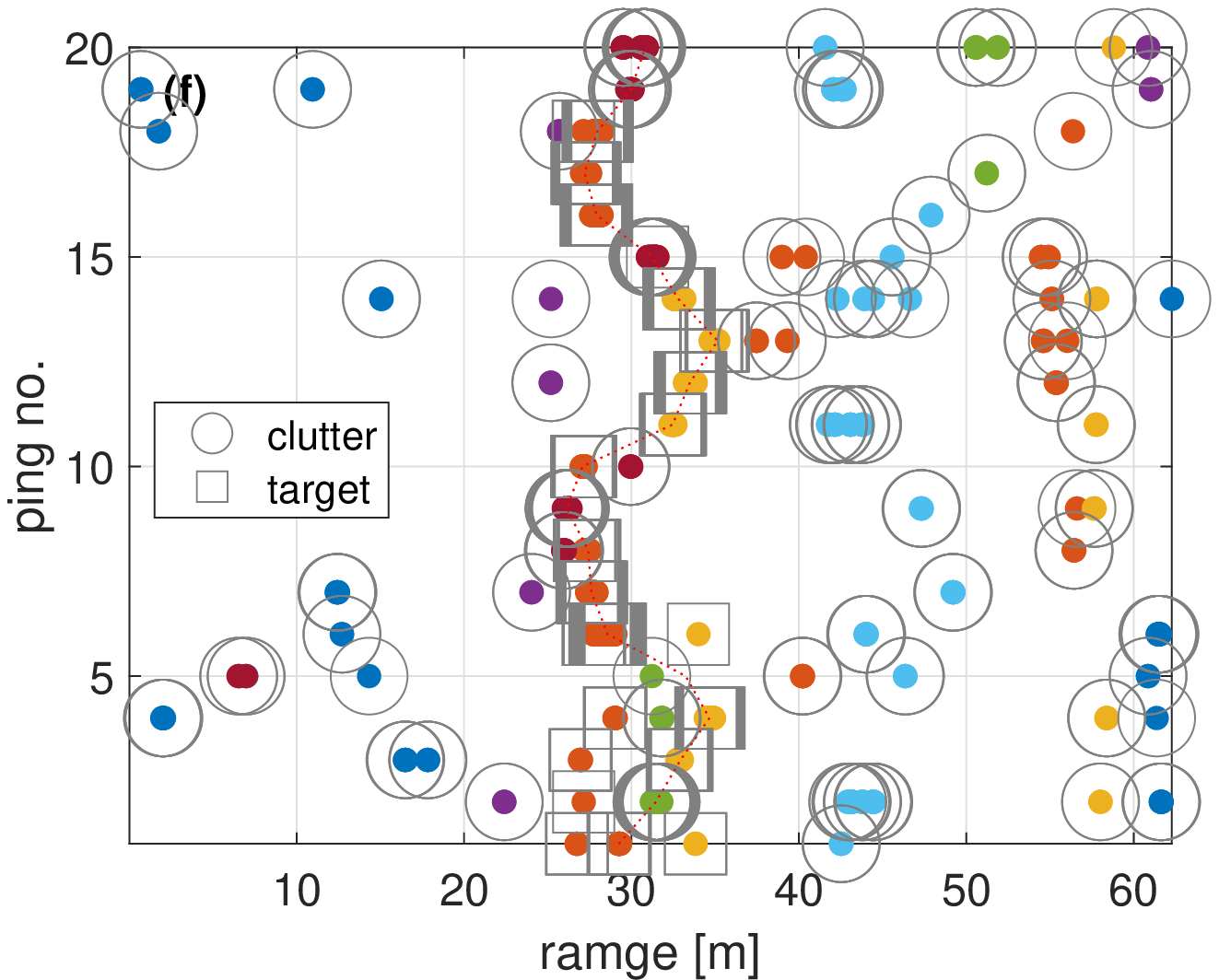}%
\label{fig_f}}
\caption{\label{fig: processing stages} Processing stages: (a) reflection matrix after MF; (b) the reflection matrix after preliminary detection; (c) graph formation; (d) choosing model order $(\epsilon=1)$; (e) clusters (color coded); (f) classification.}
\label{fig_sim}
\end{figure*}

Classification of the formed clusters to targets of interest, illustrated in Fig.~\ref{fig: processing stages}f, is based on the statistics of the data-points within the identified cluster.
The selection of features for classification is application dependent, and should reflect the difference between the desired target and clutter.
Recall that for the task of distinguishing between reflections from turtles and from clutter, we expect that:
\begin{enumerate}
	\item Reflections from a sea turtle will form clusters with relatively high connectivity, while clutter-related data-points will form clusters that are less connected.
	\item Reflections from a sea turtle will exhibit a diverse frequency spectrum, due to interference from the turtle's internal organs, while clutter-related reflections will have a more uniform frequency spectrum.
\end{enumerate}
To quantify how well a cluster is connected, we define the \textit{connectivity} of a cluster as the sum of all connections \eqref{eq: w_ij} within the cluster.
For the $k$th cluster, the connectivity is calculated by
\begin{equation} \label{eq: c_k}
	c_k = \sum_{i,j \in \mathrm{cluster}~k} w_{ij},
\end{equation}
or, in a matrix form, 
\begin{equation} \label{eq: c_k matrix form}
	c_k =  {\bf b}_k^\mathrm{T} {\bf W} {\bf b}_k \;,	
\end{equation}
where $\bf W$ holds the weights $w_{ij}$ from \eqref{eq: w_ij}.
To characterize the cluster's frequency diversity, we rank the spectral entropy of the cluster's members.
Here, a low entropy reflects high frequency diversity.
This information is obtained by performing the median,
\begin{equation}
	\bar H_k = \mathrm{median}[\{ H_i (\hat {\bf z}^{(i)}) \}_{i \in \mathrm{cluster}~k}]\;.
\end{equation}

For our application, since stability and spectral diversity are not necessarily related, we combine the two above features by
\begin{equation}
	\mathrm{class}({\bf b}_k) = 
	\begin{dcases}
		\text{target}  & (c_k > \eta_c)  \land (\bar H_k < \eta_h) \\
		\text{clutter} & ~~~~~\text{otherwise} 
	\end{dcases},\;
\end{equation}
where thresholds $\eta_c$ and $\eta_h$ are user defined.
A suggested approach to calculate these thresholds is by the distribution of features of the clutter-related data-points.
When the SCR is low, and most data-points are clutter-related, we expect the target's features to be anomalous relative to the clutter.
Since both connectivity and entropy are pre-calculated for all data-points, we can determine the thresholds based on, e.g., the empirical cumulative distribution functions of these features.
Alternatively, when the SCR is high, a large fraction of the data-points is target-related, and we cannot rely on anomaly detection.
In such cases, the clutter's entropy distribution can be estimated from a sample of the MF outputs that did not pass the threshold.
Still, to calculate connectivity, we need to apply the clustering algorithm. Consequently, we recommend learning the connectivity of the clutter offline, by using recordings from diverse sea experiments.

\subsection{Discussion}
The parameters of our above algorithm are $\alpha$, $\beta$ and $\tau$, which control the relative importance of the frequency diversity, range difference, and ping difference, respectively.
The role of these parameters is to penalize the weights of the graph's edges when pairs of data-points do not fit the model of a maneuvering target.
Setting the value too low may give low impact to the model assumptions during the clustering, while setting a value that is too high may decrease the weights, rendering the clustering ineffective.
We advise setting the above values such that $\alpha, \beta \| \Delta r_{ij} \|_{\infty}$ and $\tau \| \Delta t_{ij} \|_{\infty}$ are close to 1, and then doing some fine-tuning according to the experimental setup, e.g., by clustering together reflections from large objects, such as a ship hull or the sea floor.

Considering features for cluster classification, we note that, while the features we use are suitable for our application, one may consider additional features, such as the cluster's size and spatial extent.
Filtering out seemingly non-related data points from clusters is also possible.
However, a main source of complexity is the computation of the mutual information, which involves calculating the joint entropy of all pairs of data-points.
A suggestion to reduce this bottleneck is to split the computation of \eqref{eq: Delta_ij}, such that \eqref{eq: mutual information} will not be calculated for $\alpha << \beta \Delta r_{ij} \mathrm{u}(\Delta r_{ij}-l) + \tau \Delta t_{ij}$, where it has a negligible contribution.
Finally, we note that our solution allows for using an alternative preliminary detection method than the MF method which we presented. Examples are given in \cite{abraham2002active, yu2017fractional, berg2018classification, de2019automatic, seo2019underwater, stinco2021automatic}.


\section{Performance Analysis} \label{sec: Performance analysis}
The task we explore can be considered a mixture of detection and classification.
\textit{Detection} is considered when we wish to identify the existence/absence of a target.
We measure detection per-scenario by the trade-off between the probability of detection, $P_{\mathrm{D}}$, and the probability of false alarm, $P_{\mathrm{FA}}$.
\textit{Classification} of individual data-points to either clutter or a target is considered a generalization of detection to account for a soft threshold to the indicated existence/absence of a target.
We measure the performance of our classifier per-data-point, and evaluate it by the precision and recall trade-off.
As a benchmark, we use the Viterbi-algorithm (VA)-based detection method \cite{diamant2019active}, which performs track-before-detect to find a target's path within a clutter by probabilistic analysis.
\newline
We define a \textit{true detection} as at least one cluster per the emission of $N_p$ pings that is correctly classified as a target.
Similarly, we define a \textit{false alarm} as at least one cluster per the emission of $N_p$ pings that is falsely classified as a target.
Formally, $P_\mathrm{D}$ and $P_\mathrm{FA}$ are calculated by
\begin{subequations}\label{eq: PD and PFA}
	\begin{align}
		&  P_\mathrm{D} = \frac{
		\begin{matrix}
	        \text{number of scenarios with} \\
	        \text{targets that are detected}
	    \end{matrix}
		}{
		\begin{matrix}
	        \text{total number of simulated} \\
	        \text{scenarios without target}
	    \end{matrix}
		} \\
		& P_\mathrm{FA} = \frac{
	    \begin{matrix}
	        \text{number of scenarios without target,} \\
	        \text{but detection is determined}
	    \end{matrix}
	    }
	    {
	    \begin{matrix}
	        \text{total number of simulated} \\
	        \text{scenarios without target}
	    \end{matrix}
		}
	\end{align}
\end{subequations}
To evaluate classification, we define \textit{precision} by
\begin{equation}
	\text{precision} = \frac{ n_\mathrm{TP} }{ n_\mathrm{TP} + n_\mathrm{FP} },
\end{equation}
where $n_\mathrm{TP}$ is the number of data-points correctly classified as a target, and $n_\mathrm{FP}$ is the number of data-points falsely classified as a target.
The recall is defined by
\begin{equation}
	\text{recall} = \frac{ n_\mathrm{TP}}{ n_\mathrm{TP} + n_\mathrm{FN}},
\end{equation}
where $n_\mathrm{FN}$ is the number of data-points that are indeed a target, but are falsely classified as a non-target.
The ground truth for classification is obtained according to the distance of a determined valid data-point to the simulated path of the target, such that
\begin{equation}
	\text{ground truth class for }(r_i, p_i) = 
	\begin{dcases}
		\text{target}  & |r_i - \rho(p_i)| < \rho_0 \\
		\text{clutter} & |r_i - \rho(p_i)| \geq \rho_0 
	\end{dcases}
\end{equation}
where $\rho(p_i)$ is the simulated target's path at ping $p_i$, and $\rho_0=0.5$~m.

\subsection{Numerical Simulations}

\subsubsection{Simulation Setup}
The simulation facilitates acoustic reflections from a maneuvering target.
We use a chirp signal, $s(t)$, of a $T_\mathrm{s}=10$~ms duration and a frequency band of 7-17~kHz.	
A single simulation scenario is determined as a set for $N_\mathrm{p}=20$ pings.
The time segment duration of data for each ping is 90~ms, corresponding to a two-way range of $\sim$70~m.
For each scenario, the target's path is simulated by a random process, $\rho_{m+1} = \rho_m + n_m$, where $n_m \sim \mathrm{N}(0, \sigma_\mathrm{n}=2~\text{m}), m=1,...N_\mathrm{p}$.
We simulate the acoustic signal associated with ping $m$, $y^{(m)}_\mathrm{sim}(t)$, by
\begin{equation}
	y^{(m)}_\mathrm{sim}(t) = \\
	\begin{dcases}
		s(t-T_m) * h(t - T_m), & T_m \leq t \leq T_m+T_\mathrm{s} \\
		y_\mathrm{clutter}(t),                                               & \text{otherwise}
	\end{dcases}
\end{equation}
where $T_m=\frac{2 \rho_m}{c}$, $y_\mathrm{clutter}(t)$ are clutter samples and $h(t)$ is the simulated target's impulse response.
We consider the case where the turtle's impulse response distorts the transmitted signal $s(t)$ in a manner which we cannot predict. We therefore simulate $h(t)$ by a multivariate random variable, ${\bf h}_{100 \times 1}$, that is uniformly distributed in [-1  1], and then normalized such that ${\bf h}^\mathrm T {\bf h} = 1$.
The same impulse response is used for all pings within a single simulation scenario.
To explore performance over a realistic scenario, clutter samples $y_\mathrm{clutter}(t)$ are taken from sea experiments. In these experiments, we deployed our sonar to a depth of roughly 20~m, and transmitted the same chirp signals used in the simulation.
The bottom depth was 125~m, and a manual inspection ensured that the reflection pattern contained only clutter.
The overall duration of obtained clutter samples is 257~s, from which we randomly uniformly choose in each simulation run.
The desired SCR is achieved by controlling the amplitude of the simulated target's echo.
No ambient noise is added.
We run $300$ simulation scenarios per SCR value including target, and $1000$ scenarios to explore false alarms.

\subsubsection{Simulation Results}
\begin{figure}[h]
	\centering
	\includegraphics[width=0.4\textwidth]{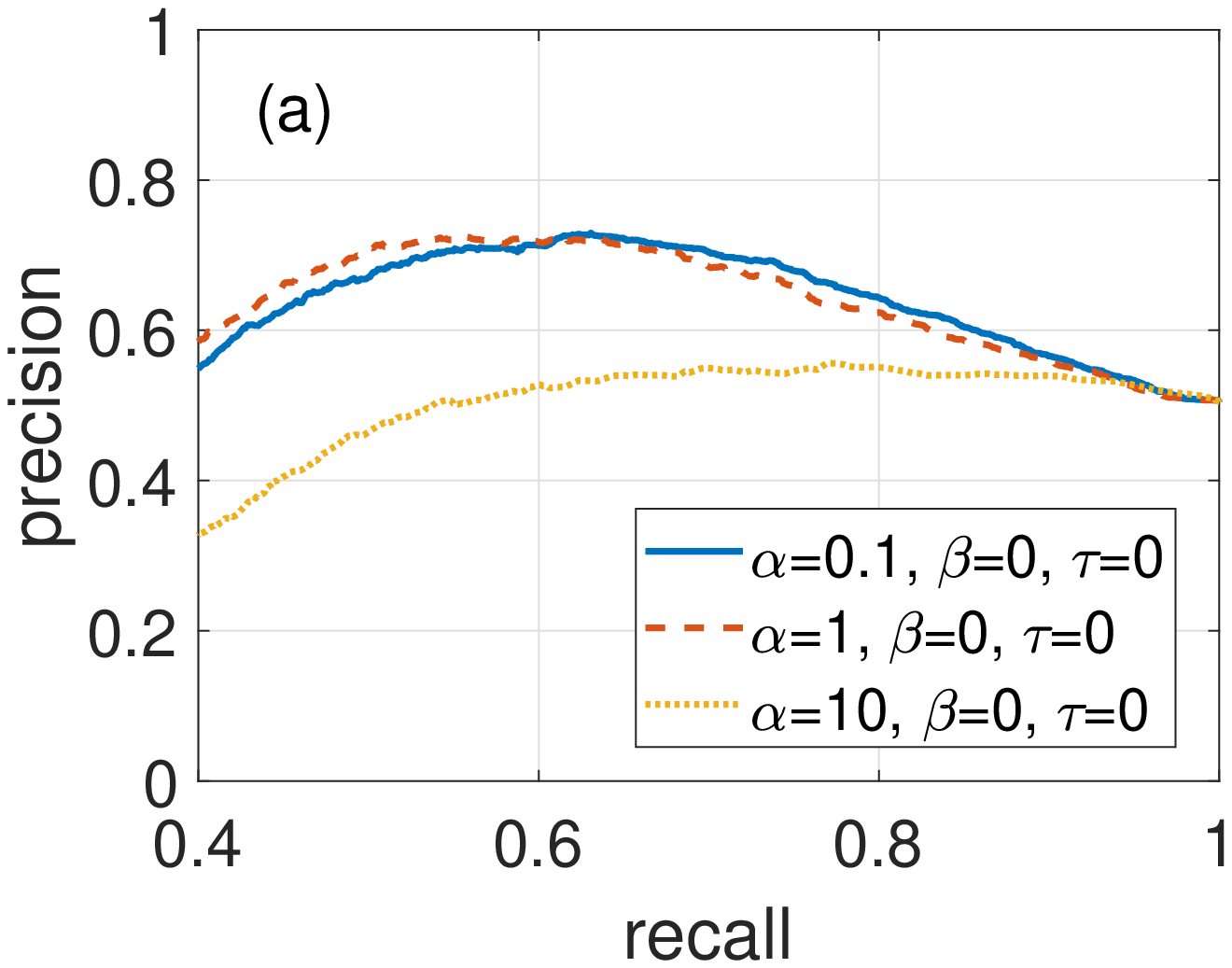}
	\includegraphics[width=0.4\textwidth]{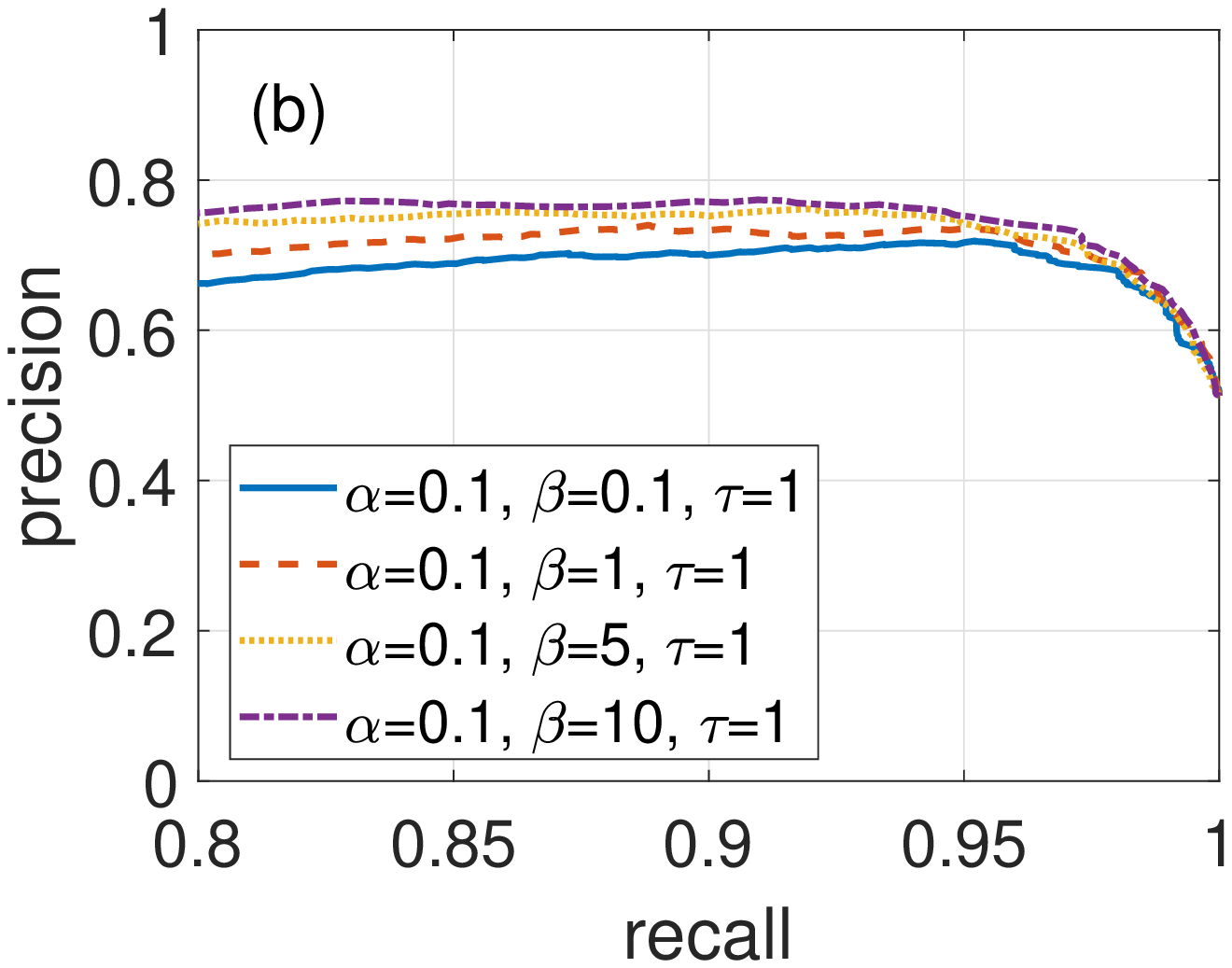}
	\includegraphics[width=0.4\textwidth]{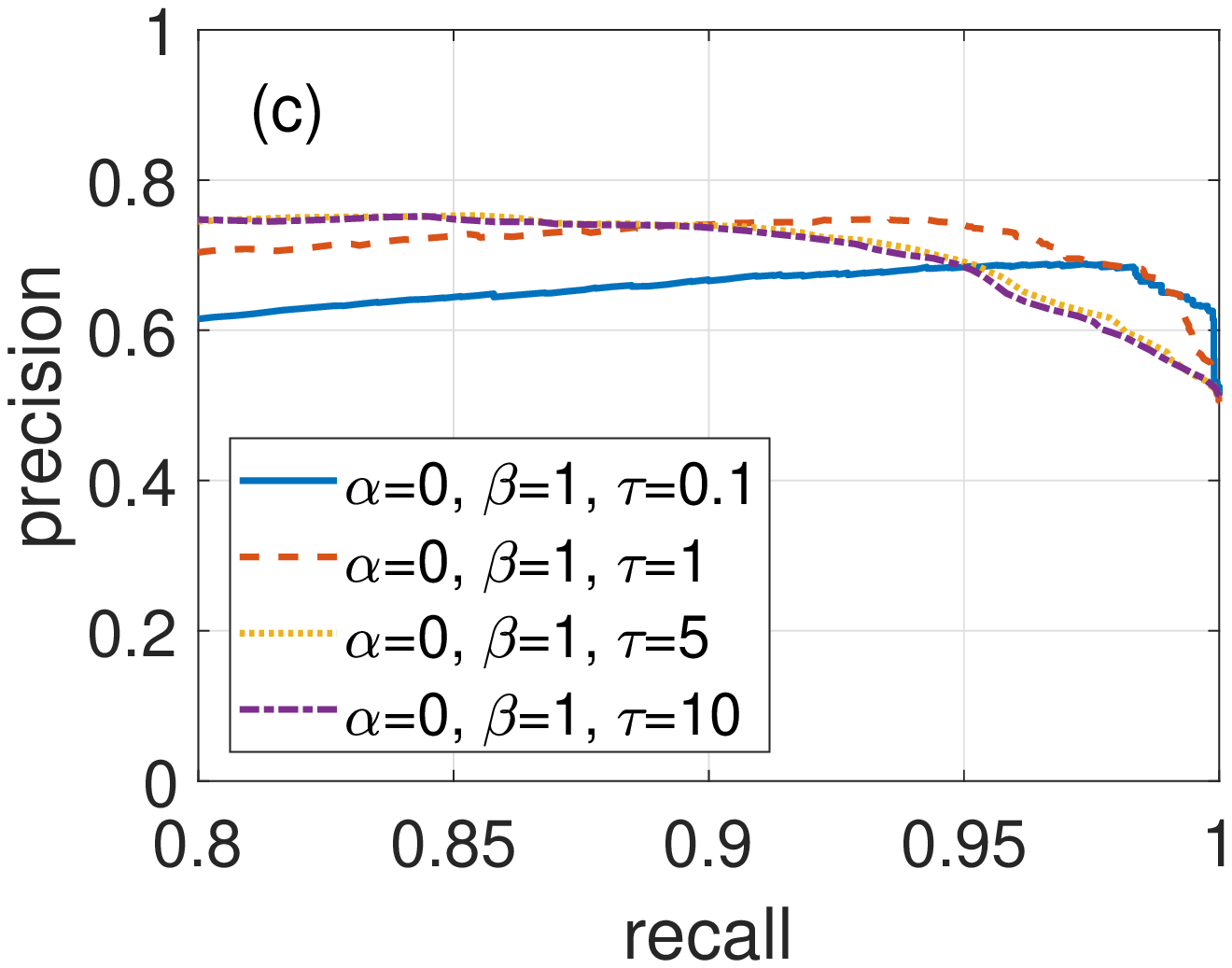}
	\caption{ \label{fig: parameters_sensitivity} Precision-recall curves for exploring sensitivity to parameters, while changing the connectivity threshold. SCR = -9dB. (a) sensitivity to $\alpha$, (b) sensitivity to $\beta$, (c) sensitivity to $\tau$.}
\end{figure}

We start by exploring the sensitivity of our algorithm for parameters $\alpha, \beta$ and $\tau$.
The classification sensitivity for $\alpha$ is presented in Fig.~\ref{fig: parameters_sensitivity}a, while setting $(\beta, \tau) = 0$, and changing the threshold $\eta_\mathrm{c}$.
The observable maxima in the precision-recall curve indicates that there is a connectivity threshold $\eta_\mathrm{c}$ under which detection degrades drastically.
From this figure, it is evident that the frequency diversity alone, without using the spatial and temporal information, contains sufficient information.
We also observe that classification performance is sensitive to $\alpha$.
This is because the exponent in \eqref{eq: w_ij} is multiplied by $\alpha$ and thus any change in $\alpha$ has an impact on weights $w_{ij}$.
A similar trend is observed in Fig.~\ref{fig: parameters_sensitivity}b, which shows the classification results when only $\beta$ varies, i.e., exploring the impact of the spatial feature in the reflections on classification. Here, we set $\alpha=0$, but $\tau=1$, since we need the temporal connection to form clusters that are geometrically valid.
We observe that both precision and recall improve with $\beta$, but converge for $\beta>5$.
Next, the sensitivity for the temporal feature in the reflections is explored by changing parameter $\tau$, keeping $\beta=1$ and $\alpha=0$.
The results in Fig.~\ref{fig: parameters_sensitivity}c show a trade-off: a lower $\tau$ increases the recall at the cost of precision, i.e., the algorithm tends to classify too many instances as a 'target'.
On the other hand, a higher $\tau$ reduces both false-positives and true-positives.
It seems that a favorable trade-off is $\tau=1$.
In the following, we pick the values $\alpha=0.1$, $\beta=1$, and $\tau=1$.

\begin{figure}[h]
	\centering
	\includegraphics[width=0.4\textwidth]{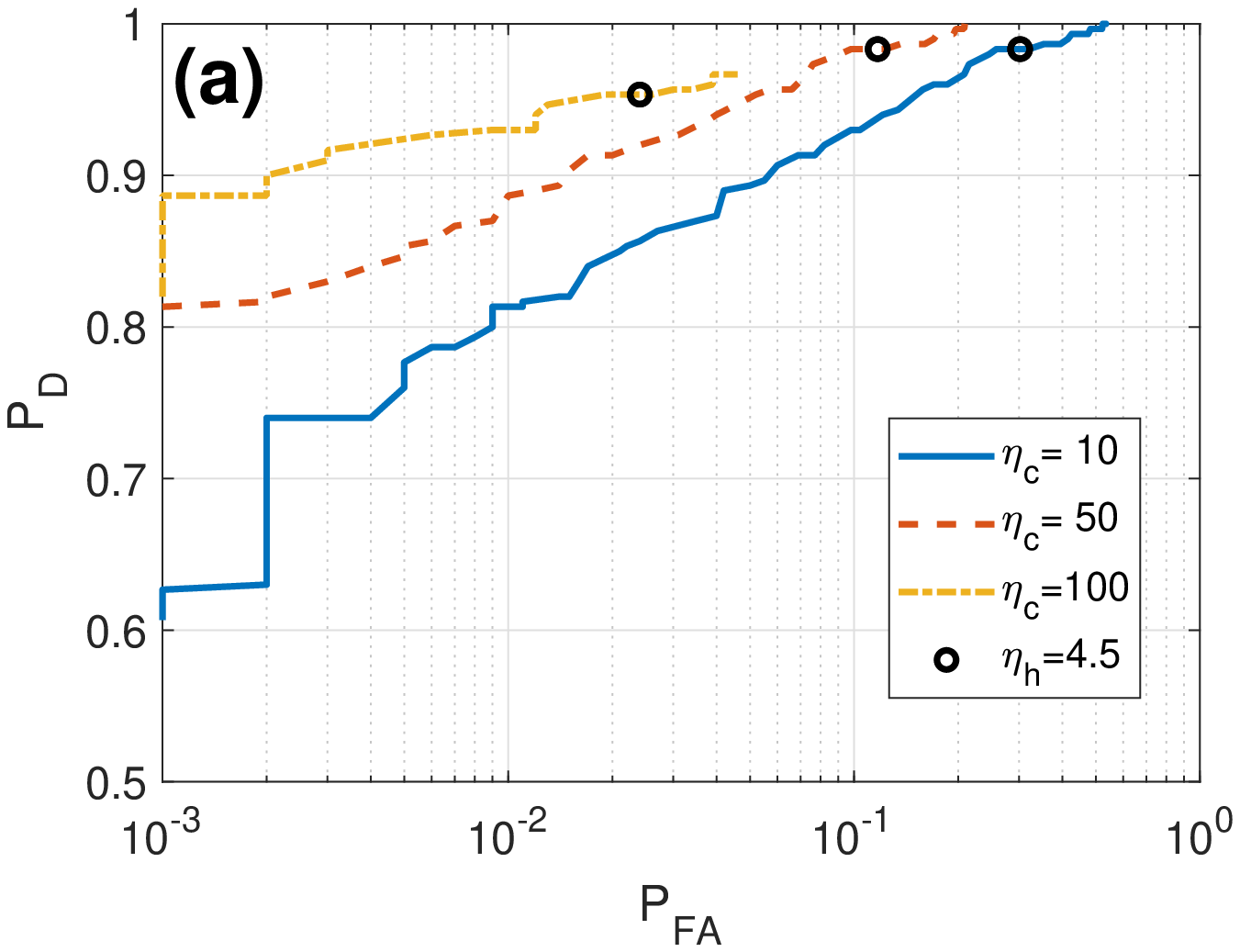}
	\includegraphics[width=0.4\textwidth]{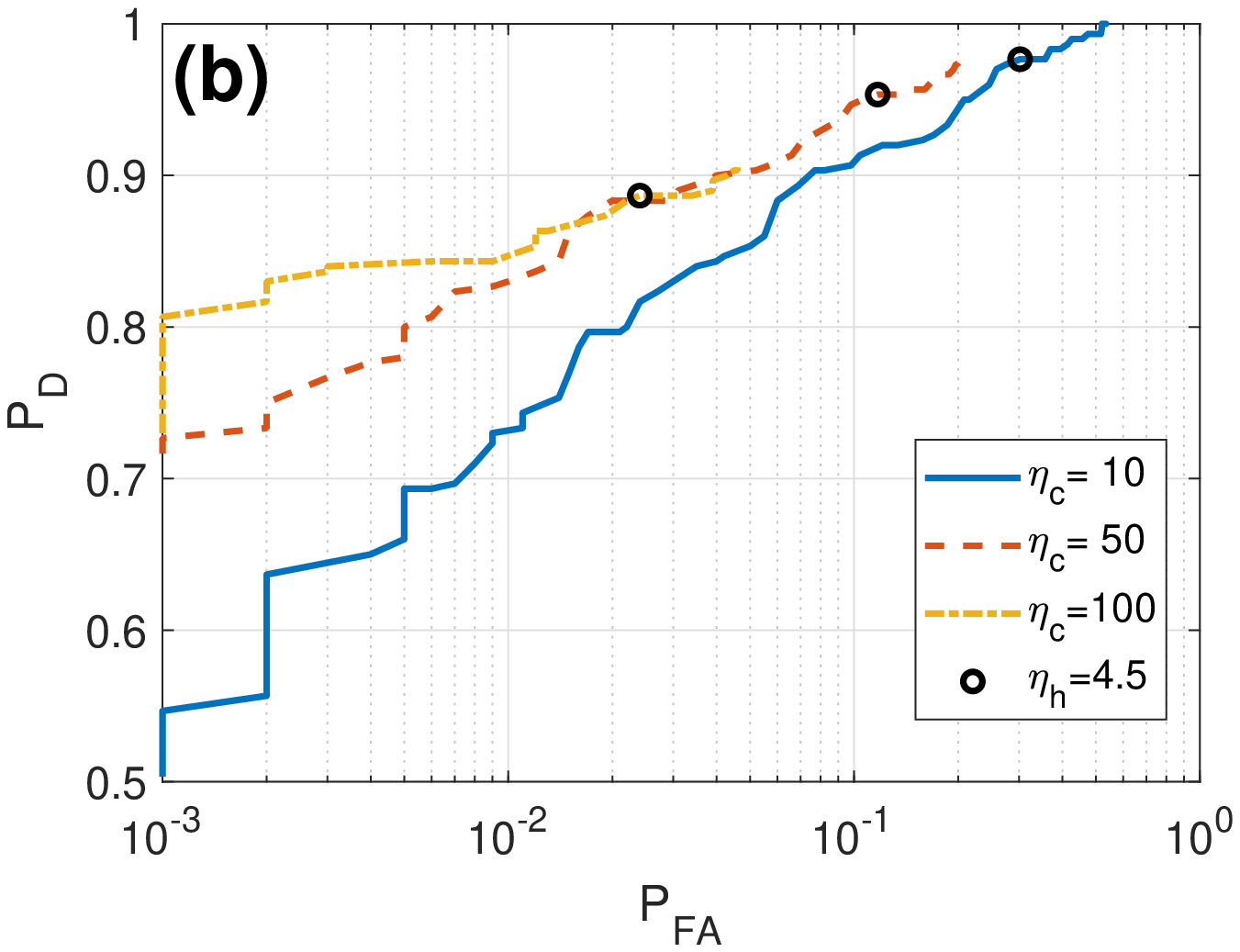}
	\caption{ \label{fig: entropy_ROC} ROC where $\eta_\mathrm{h}$ is parameter, and $(\alpha, \beta, \tau) = (0.1, 1, 1)$. Percentage of valid pings is (a) 100\%, (b) 70\%. SCR = -9dB.}
\end{figure}

The effect of the entropy threshold is explored in Fig.~\ref{fig: entropy_ROC}, where $\eta_\mathrm{h}$ is used as a parameter, with $2 \leq \eta_\mathrm{h} \leq5$, while three values for $\eta_\mathrm{c}$ are tested.
Here, the SCR is -9~dB.
Fig.~\ref{fig: entropy_ROC}a shows results for the case where the reflections of each of the $N_\mathrm{p}=20$ emissions include a valid target in addition to clutter.
In contrast, Fig.~\ref{fig: entropy_ROC}b shows the case where only 70\% of the 20 emissions include a valid reflection, such that the target appears as 'blobs' of sporadic detections.
We observe that the performance of our algorithm degrades in this latter case, since fewer emissions are available for detection.
In both cases, it is evident that higher $\eta_\mathrm{c}$ produces better ROCs, but the maximal $P_\mathrm{D}$ that is reached is limited.
This is because target-related clusters with lower connectivity are misclassified as clutter.
The circles markers on the plots are for $\eta_\mathrm{h}=4.5$, which we identify as a reasonable trade-off that keeps $P_\mathrm{D}$ relatively constant for a large range of $\eta_\mathrm{c}$ values.

\begin{figure}[h]
	\centering
	\includegraphics[width=0.35\textwidth]{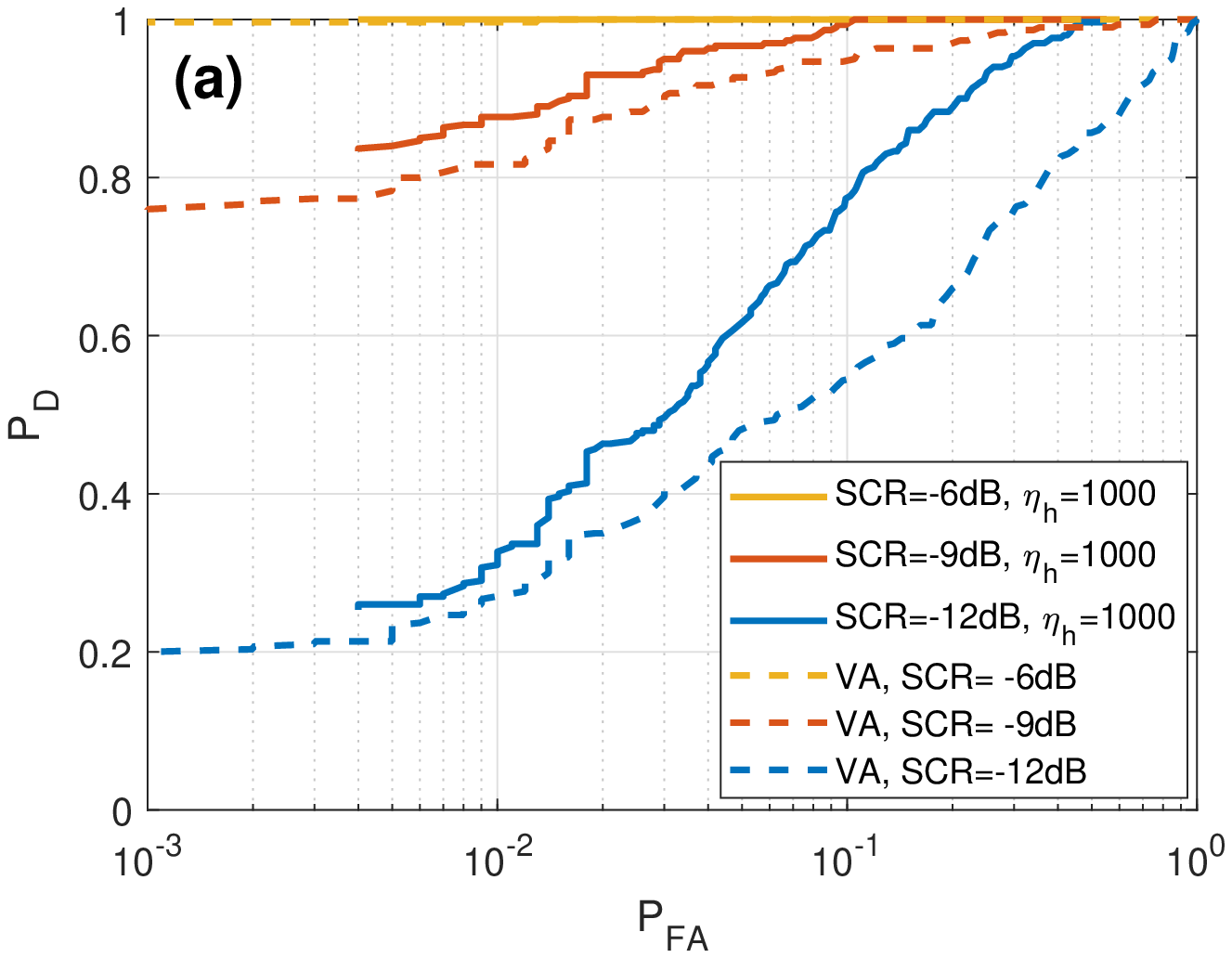}
	\includegraphics[width=0.35\textwidth]{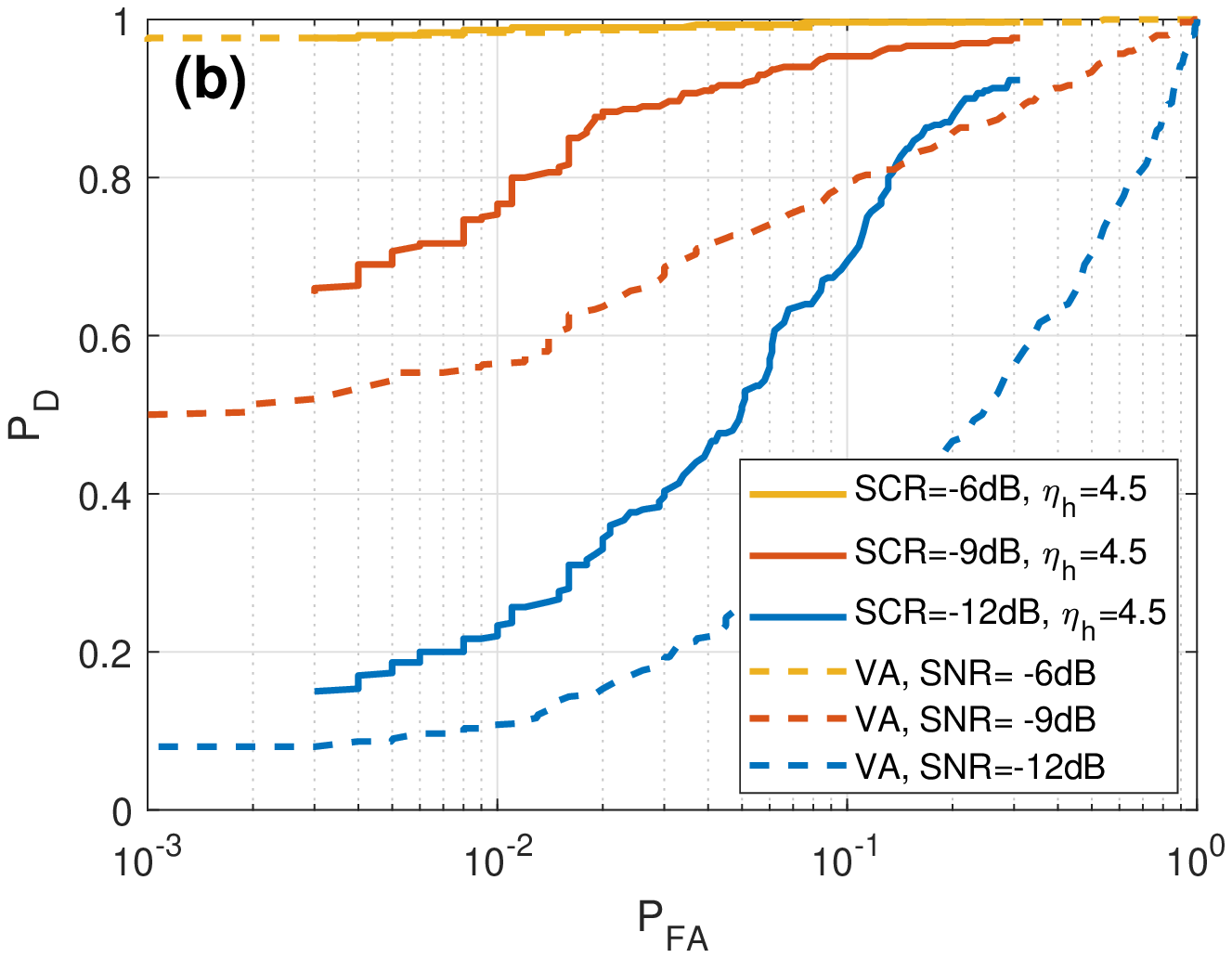}
	\caption{ \label{fig: ROC_connectivity} ROC results when: (a) all emissions include a valid reflection, (b) $70\%$ of the emissions are valid. $(\alpha, \beta, \tau) = (0.1, 1, 1)$.}
\end{figure}

Fig.~\ref{fig: ROC_connectivity} shows the impact of the SCR level on the detection performance, and the two cases of 100\% and 70\% valid targets in the 20 emissions are presented in Figs.~\ref{fig: ROC_connectivity}a,b, respectively.
Here, we set the entropy threshold to be $\eta_\mathrm{h}=4.5$.
Additional fixed parameters are $\eta_\mathrm{MF}=5 \cdot 10^{-6}$, $(\alpha, \beta, \tau) = (0.1, 1, 1)$.
The track-before-detect VA benchmark is marked by a dashed line.
As expected, both algorithms perform well for high SCR=-6~dB, while performance degrades rapidly as the SCR decreases.
It is evident that, for low SCR, our algorithm outperforms the benchmark.
Additionally, in the case of 70\% valid pings, the performance of the VA-based approach highly degrades, while our method, although affected, is more robust.
This is because our method is designed to look for 'blobs' of detection, while the VA-based method assumes that the reflections form a path.

\subsection{Field Experiments}
\begin{figure}[h]
	\centering
	\includegraphics[width=0.4\textwidth]{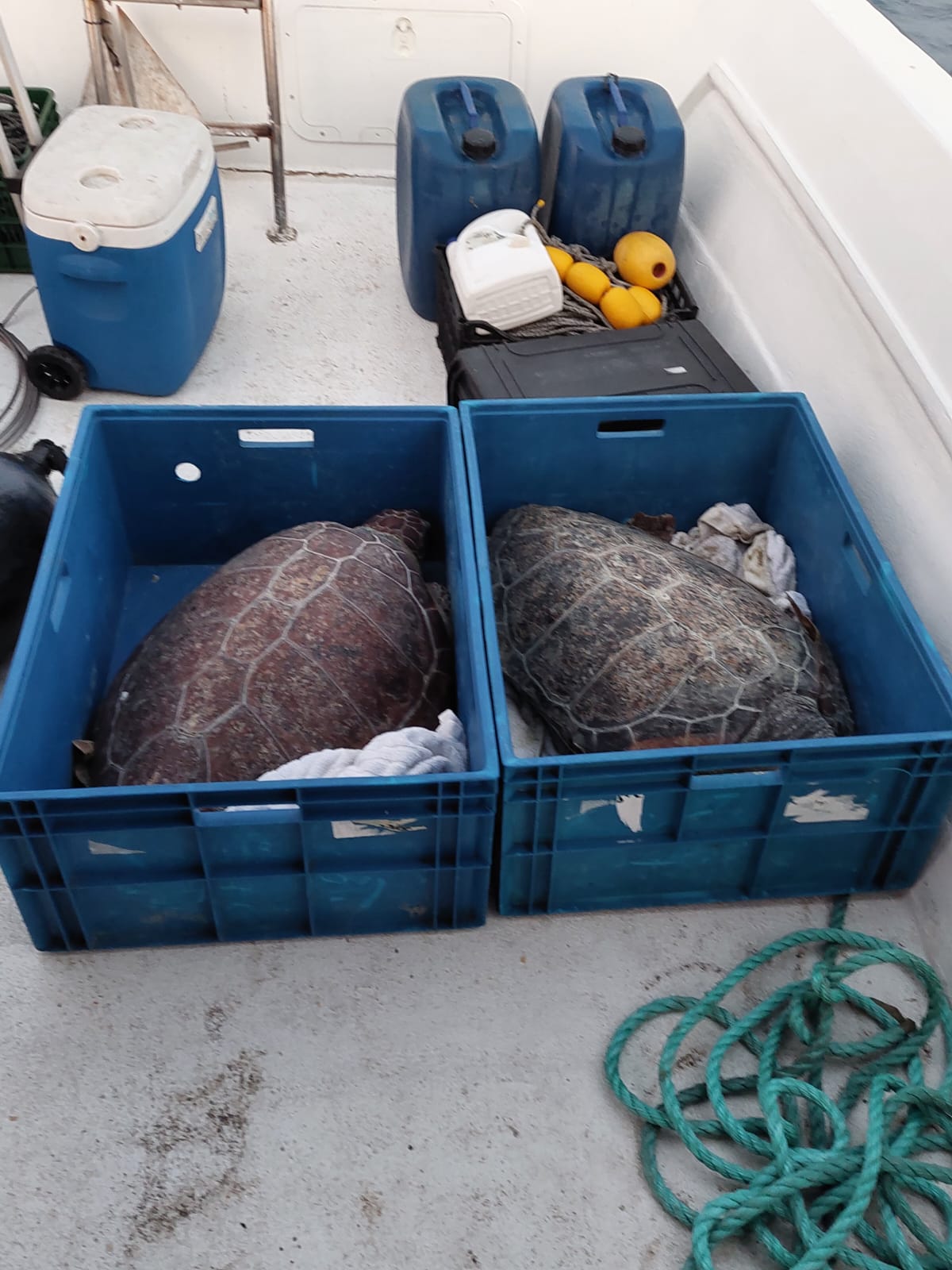}
	\includegraphics[width=0.4\textwidth]{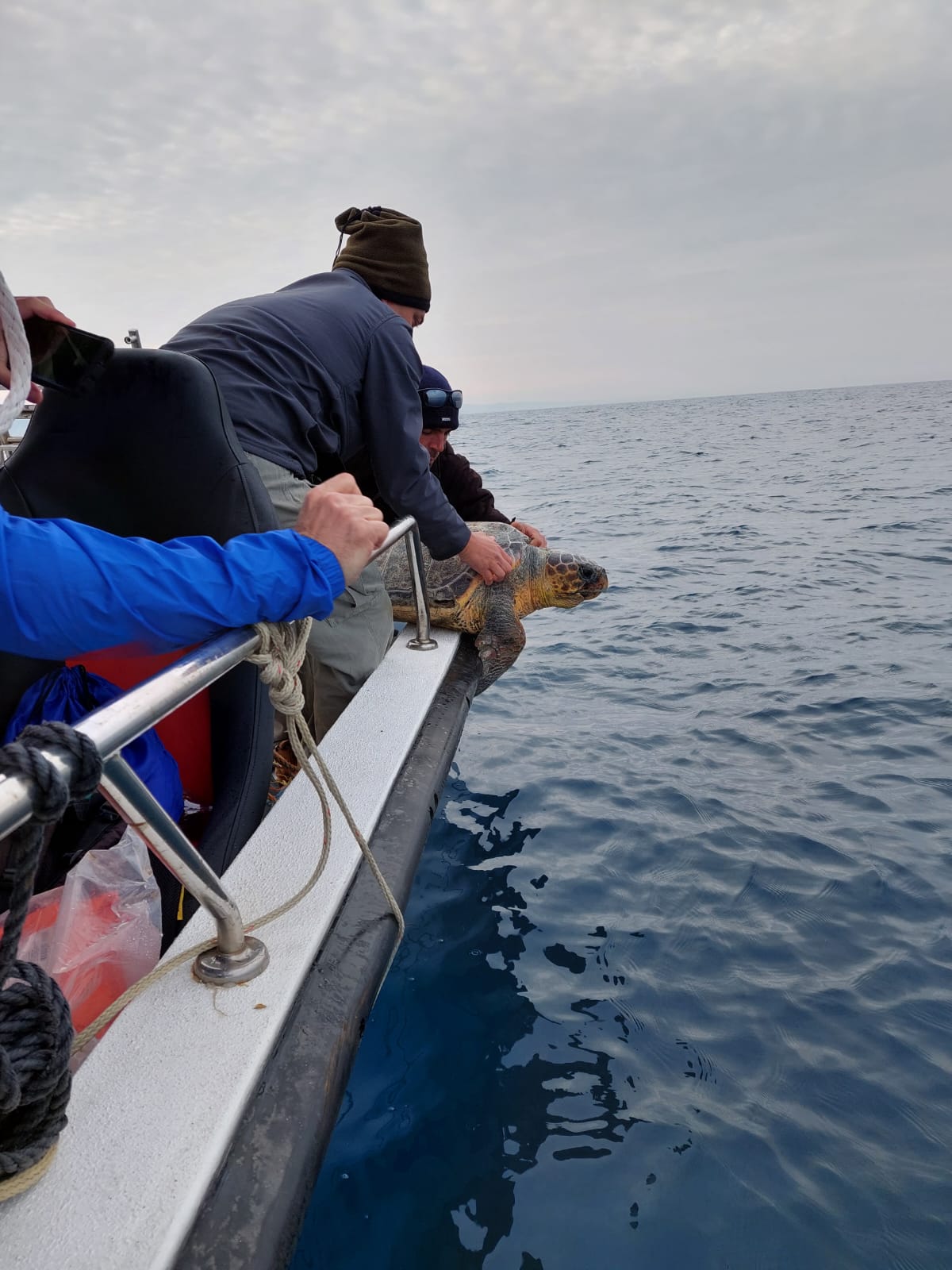}
	\caption{ \label{fig: turtles} The sea turtles that were released, and one of the release activities.}
\end{figure}

The above simulation results explored the performance statistically, but lack a reliable model for the acoustic reflections from the sea turtle.
To this end, we conducted a designated sea experiment that involved the release of two rehabilitated sea turtles, and measured their acoustic reflections.
The experiment was carried out roughly 3~km off the coast of Haifa, Israel, in 30~m of water.
The sea turtles' released weight was roughly 40~kg, and their size was roughly 60~cm in diameter.
These were mature turtles that had suffered a shock wave and were rehabilitated in the Israeli Rescue Center for Sea Turtles. The release was performed by the center's personnel, under the approval of the Israeli Nature and Park Authority (NPA).
An image of the release of one of the turtles is shown in Fig.~\ref{fig: turtles}.

\subsection{Experimental Setup}
\begin{figure}[h]
	\centering
	\includegraphics[width=0.48\textwidth]{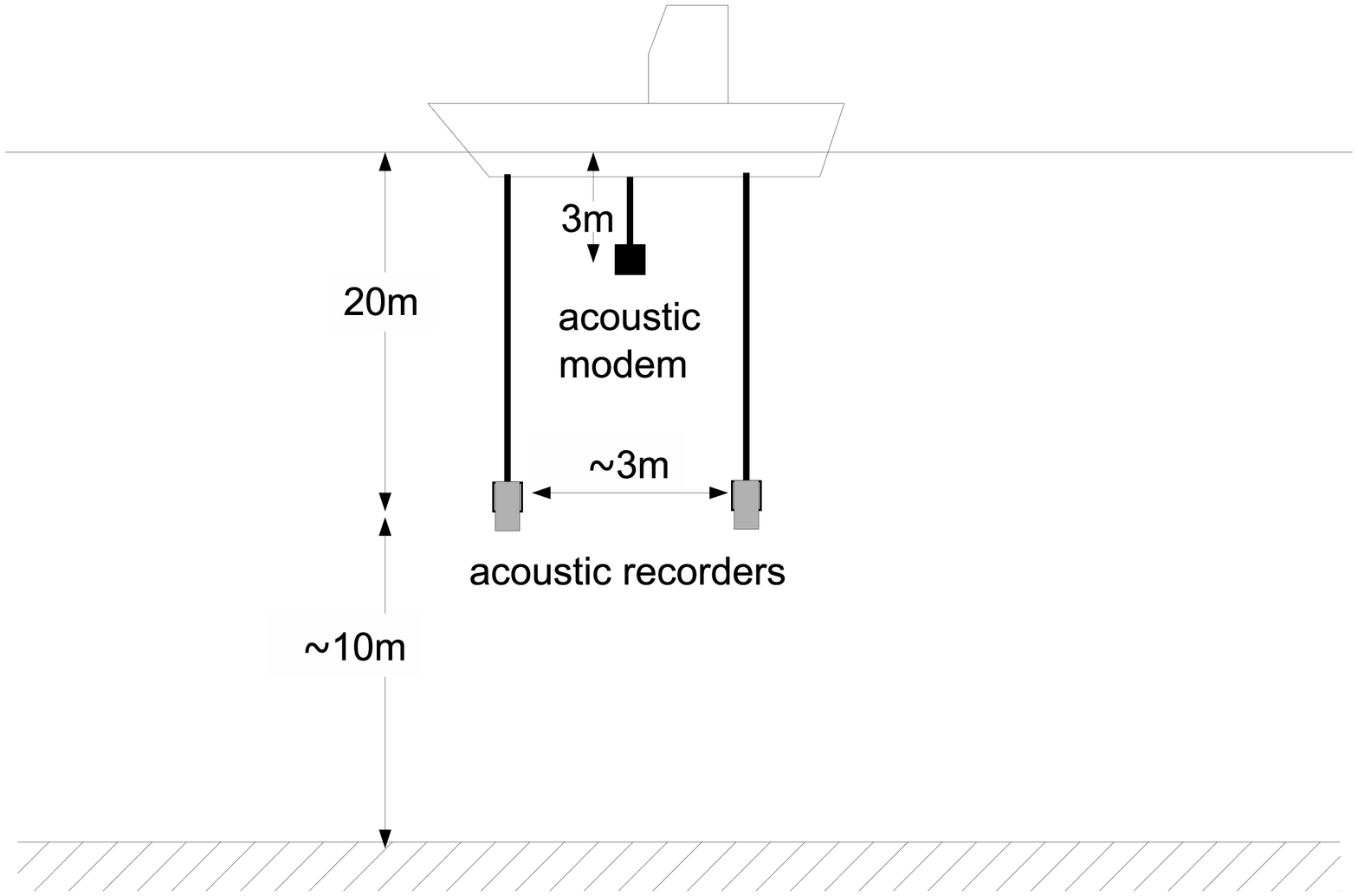}	
	\caption{\label{fig: experimental setup}The sea experiment setup.}
\end{figure}

The experimental setup is illustrated in Fig.~\ref{fig: experimental setup}.
To transmit the chirp signals, we used the Evologics S2CR~7/17 and S2C~M~48/78 software-defined acoustic modems to emit trains of chirp signals of 10~ms duration at a PRI of 0.7~s and at frequency ranges of 7-17~kHz and 48-78~kHz, respectively. The emission power was 170~dB re. $1 \mu$P@1m.
The modems were deployed from the boat to a depth of 3~m.
Two omni-directional self-made receivers were deployed from the boat to a depth of 20~m.
Continuous recording took place at 192k samples per second with 3 Byte resolution per sample.
The modems started emitting the signals prior to the release of the sea turtles.
During these events, the turtles mostly swam on the surface, but also submerged from time to time. We visually observed their behavior from the vessel, and documented the rise and submerge times.

\subsection{Experimental Results}

\subsubsection{Detection of a sea turtle}
\begin{figure}[h]
	\centering
	\includegraphics[width=0.48\textwidth]{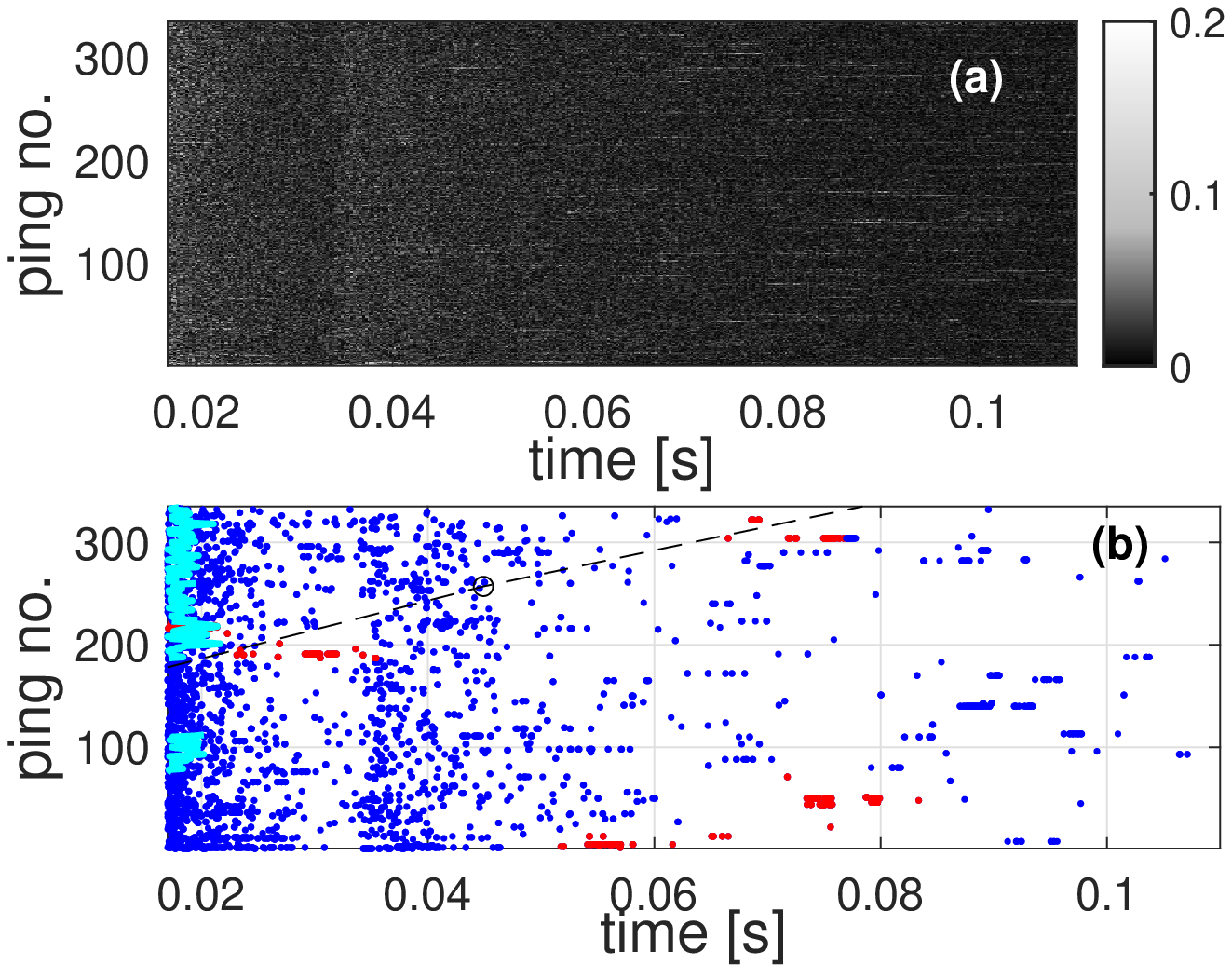}
	\includegraphics[width=0.48\textwidth]{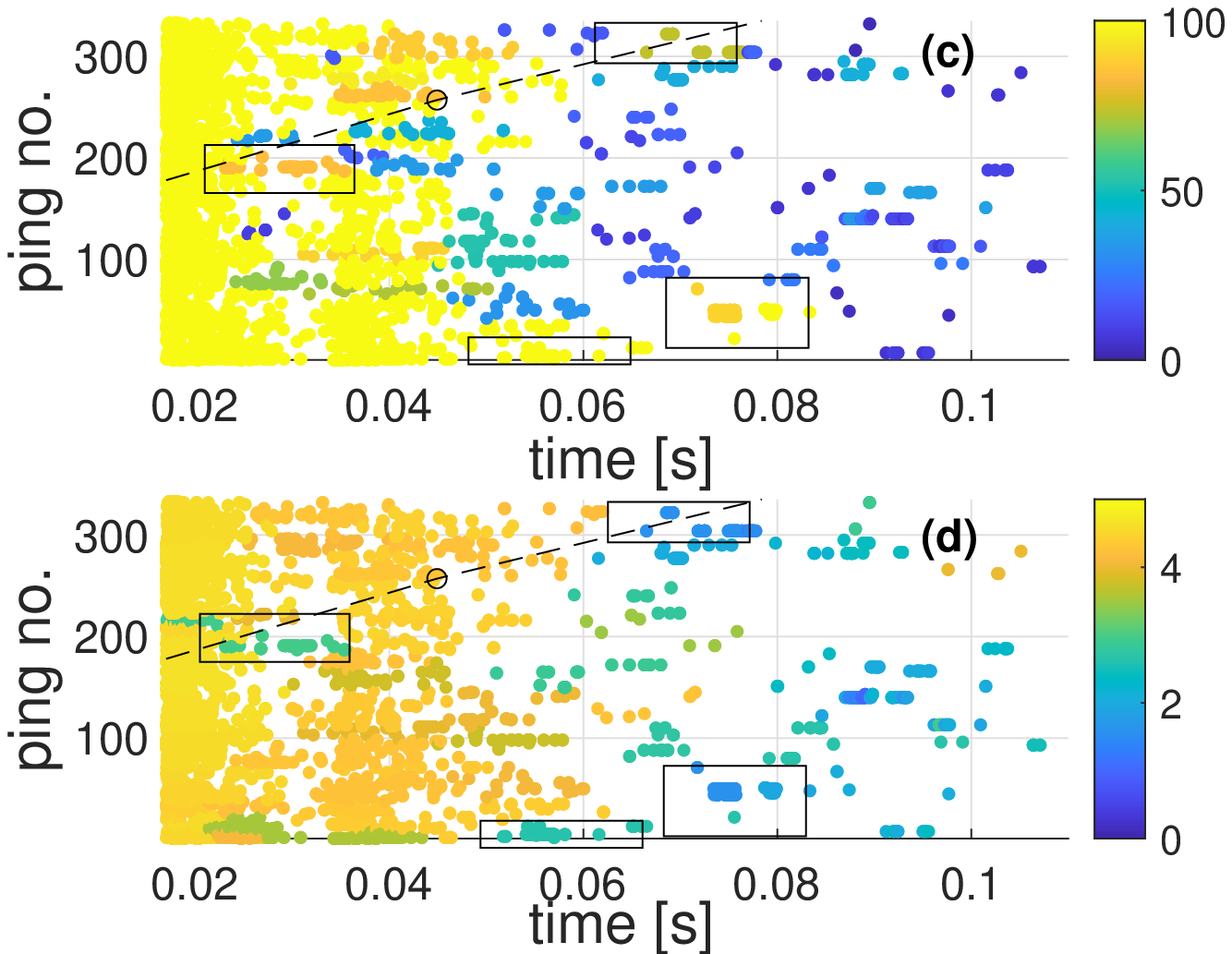}
	\caption{ \label{fig: turtle 7-17kHz results} Results for transmission signals in the range 7-17kHz. (a) Intensity  image of the MF's output. (b) Detections after threshold (our target indications are in red, VA detected path in cyan). (c) Connectivity of clusters. (d) Median entropy of clusters. Target indications are marked by rectangles in (c) and (d).}
\end{figure}

For analysis, we considered the parameter set $(\alpha, \beta, \tau) = (0.1, 1, 1)$.
We applied our algorithm on patches of 37 pings with a duration of 60~ms each.
Fig.~\ref{fig: turtle 7-17kHz results} shows results for the frequency range of 7-17~kHz.
In this experiment, there was a plastic buoy in the water with a GPS receiver at a range of no more than 20~m from the sea turtle.
The path of this GPS is marked by a dashed line.
Panel (a) shows the data-sample matrix after the MF processing, but before the thresholding operation.
Here, we merge the reflections from all the emissions.
We observe that the main reflection comes from the bottom at roughly 35~ms. 
The detection indications after thresholding are shown in Panel (b), and detections that are classified as targets are marked in red.
The path detected by the VA benchmark is plotted in cyan, using a threshold value of 94.
Panels (c) and (d) show the connectivity and the median entropy of the clusters, respectively; and clusters classified as targets are marked by rectangles.
We observe two target indications along the path of the GPS: one at emission 191 that is $\sim$ 7~m from the path, and the other at emission 304 that is right on the path.
The range to these targets, measured on the surface, is $28.7 \pm 2$~m and $60.2 \pm 4.8$~m, respectively.
In addition, there are two clusters that were falsely classified as targets on emissions 5 and 44 at ranges of $47.1 \pm 4.2$~m and $61.5 \pm 1.5$~m, respectively.
We see that the benchmark VA detects intense reflections at the leftmost part of the data-sample matrix, and follows them instead of the assumed turtle.
This is because the VA uses only intensities as an input, and is not able to detect more than a single target per-patch.
Instead, our method has the following advantages: it uses spectral information, and multiple targets can be detected.

\begin{figure}[h]
	\centering
	\includegraphics[width=0.48\textwidth]{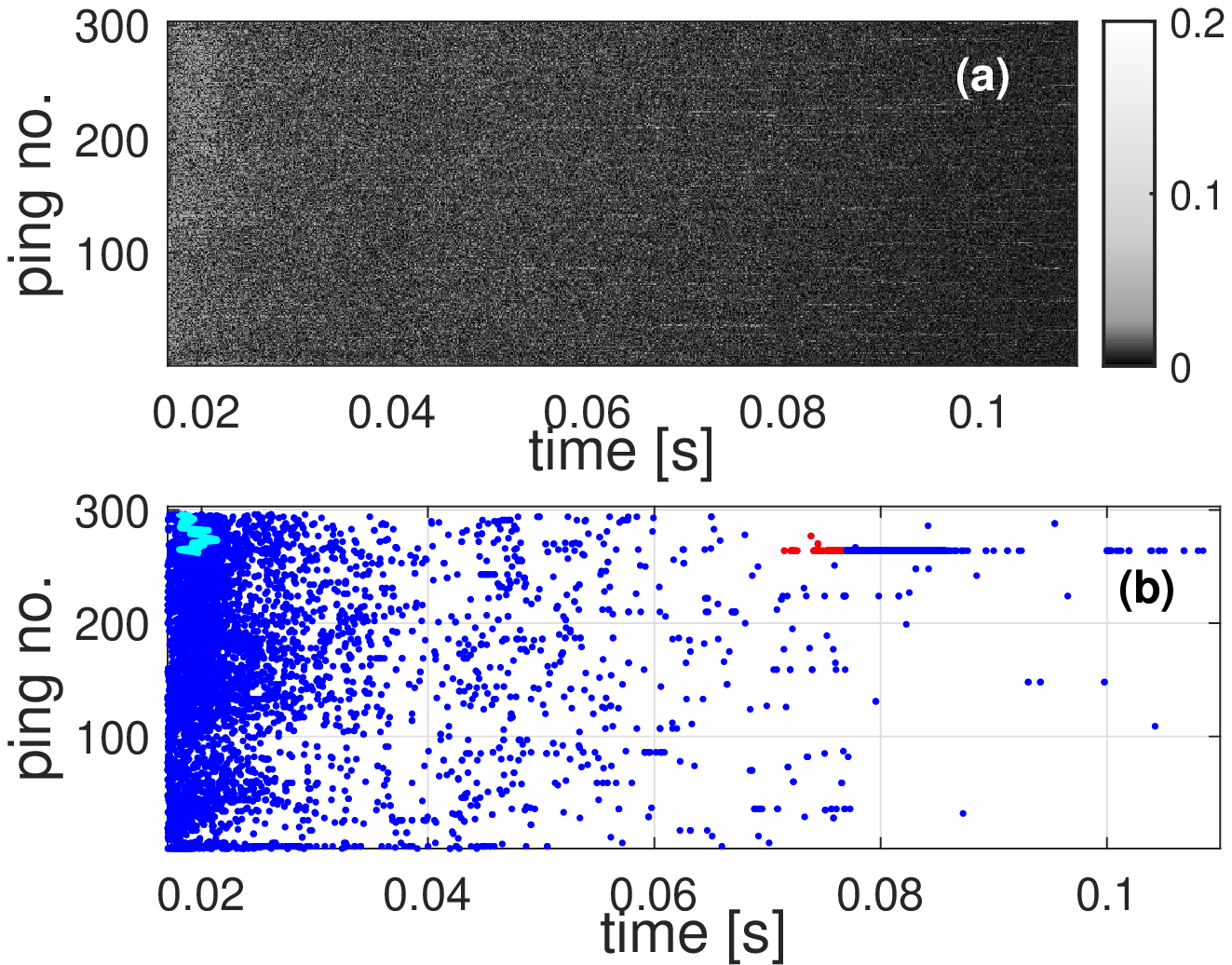}
	\includegraphics[width=0.48\textwidth]{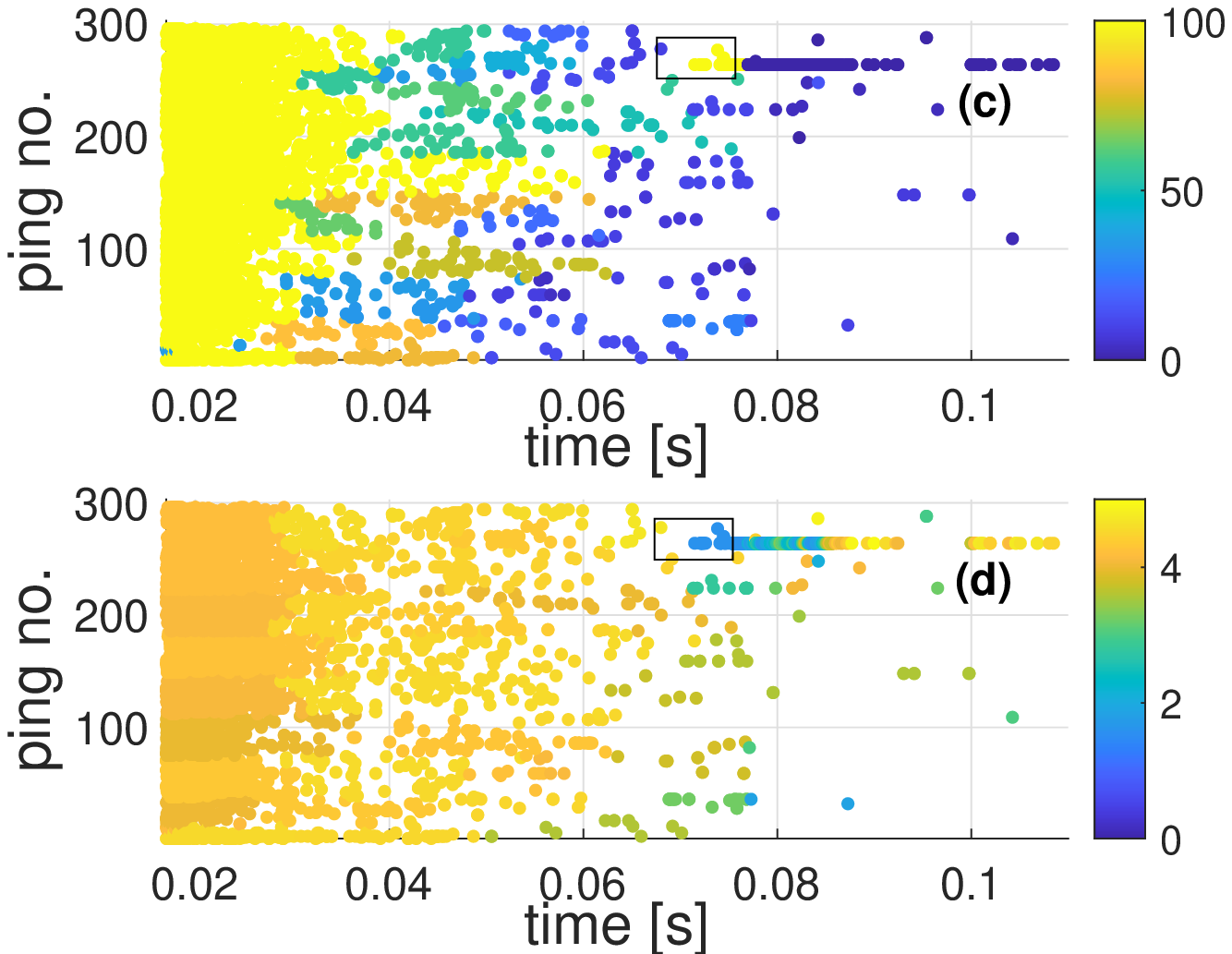}
	\caption{ \label{fig: turtle 48-78kHz results} Results for transmission signals in the range 48-78kHz.  (a) Intensity  image of the MF's output. (b) Detections after threshold (our target indications are in red, VA detected path in cyan). (c) Connectivity of clusters. (d) Median entropy of clusters. Target indications are marked by rectangles in (c) and (d).}
\end{figure}

Results obtained for the 48-78~kHz frequency range are shown in Fig.~\ref{fig: turtle 48-78kHz results}.
In this experiment, there was no GPS buoy, and we rely on the experiment's log to identify the events.
The sea turtle was released around ping 108, and was swimming on the surface, thus making its detection very challenging.
Approximately 90 seconds after its release, at emission ping 236, the turtle was observed taking air, indicating that it was about to dive.
This arguable dive enabled its detection, and indeed, we observed a detection that matches a low-entropy cluster around emission 264 with a time delay of $\sim 74$~ms, corresponding to a range of $61.9 \pm 2.1$~m.
In this case, no false alarms were observed.
Moreover, in this case, the VA benchmark, applied with a threshold of 147, is blinded by high-intensity reflections, and cannot detect the turtle.


\section{Conclusion} \label{sec: Conclusion}
In this work, we proposed an unsupervised remote sensing algorithm to identify a sea turtle within a point cloud of sonar's reflections under low SCR conditions.
We designed a clustering algorithm, which is based on the geometrical relations between a sample within the point cloud that relates to the target's reflections, as well as on the expected spectral diversity of the turtle's reflections.
We also utilized the expected stability of target-related reflections as another means of detection verification.
We explored the performance of our algorithm in both simulations and in a designated sea experiment, where we released two rehabilitated sea turtles and tracked their trajectory acoustically.
Compared to a track-before-detect benchmark, our results show an improved robustness to fluctuating intensity, and a better trade-off between the false alarm and detection rates. 
Future research may improve classification by adding more features and exploring a multivariate feature combination.


\bibliographystyle{IEEEtran}

\begin{thebibliography}{10}
	\providecommand{\url}[1]{#1}
	\csname url@samestyle\endcsname
	\providecommand{\newblock}{\relax}
	\providecommand{\bibinfo}[2]{#2}
	\providecommand{\BIBentrySTDinterwordspacing}{\spaceskip=0pt\relax}
	\providecommand{\BIBentryALTinterwordstretchfactor}{4}
	\providecommand{\BIBentryALTinterwordspacing}{\spaceskip=\fontdimen2\font plus
		\BIBentryALTinterwordstretchfactor\fontdimen3\font minus
		\fontdimen4\font\relax}
	\providecommand{\BIBforeignlanguage}[2]{{%
			\expandafter\ifx\csname l@#1\endcsname\relax
			\typeout{** WARNING: IEEEtran.bst: No hyphenation pattern has been}%
			\typeout{** loaded for the language `#1'. Using the pattern for}%
			\typeout{** the default language instead.}%
			\else
			\language=\csname l@#1\endcsname
			\fi
			#2}}
	\providecommand{\BIBdecl}{\relax}
	\BIBdecl
	
	\bibitem{nelms2016seismic}
	S.~E. Nelms, W.~E. Piniak, C.~R. Weir, and B.~J. Godley, ``Seismic surveys and
	marine turtles: An underestimated global threat?'' \emph{Biological
		conservation}, vol. 193, pp. 49--65, 2016.
	
	\bibitem{levy2015small}
	Y.~Levy, O.~Frid, A.~Weinberger, R.~Sade, Y.~Adam, U.~Kandanyan, V.~Berkun,
	N.~Perry, D.~Edelist, M.~Goren \emph{et~al.}, ``A small fishery with a high
	impact on sea turtle populations in the eastern mediterranean,''
	\emph{Zoology in the Middle East}, vol.~61, no.~4, pp. 300--317, 2015.
	
	\bibitem{macdonald2012home}
	B.~D. MacDonald, R.~L. Lewison, S.~V. Madrak, J.~A. Seminoff, and T.~Eguchi,
	``Home ranges of east pacific green turtles \textit{{C}helonia mydas} in a
	highly urbanized temperate foraging ground,'' \emph{Marine Ecology Progress
		Series}, vol. 461, pp. 211--221, 2012.
	
	\bibitem{bolten1994seasonal}
	A.~B. Bolten, \emph{Seasonal abundance, size distribution, and blood
		biochemical values of {L}oggerheads (\textit{Caretta caretta}) in Port
		Canaveral Ship Channel, Florida}.\hskip 1em plus 0.5em minus 0.4em\relax US
	Department of Commerce, National Oceanographic and Atmospheric
	Administration, National Marine Fisheries Service, Southeast Fisheries
	Science Center, 1994, vol. 353.
	
	\bibitem{mahfurdz2018green}
	A.~Mahfurdz, S.~Saifullah \emph{et~al.}, ``Green turtle and fish identification
	based on acoustic target strength,'' \emph{International Journal of Advances
		in Intelligent Informatics}, vol.~4, no.~1, pp. 53--62, 2018.
	
	\bibitem{perez2013ts}
	I.~P{\'e}rez-Arjona, V.~Espinosa, E.~Alonso, P.~Ord{\'o}{\~n}ez, S.~Llorens,
	V.~Puig, M.~Rodilla, J.~Casta{\~n}o, J.~Esteban, and J.~Eymar, ``{TS}
	measurements and simulations of {M}editerranean sea turtles,'' in
	\emph{Proceedings of the 38th Scandinavian Symposium on Physical Acoustics,
		Geilo, Norway}, 2013, pp. 1--6.
	
	\bibitem{jones2017broadband}
	B.~A. Jones, T.~K. Stanton, J.~A. Colosi, R.~C. Gauss, J.~M. Fialkowski, and
	J.~Michael~Jech, ``Broadband classification and statistics of echoes from
	aggregations of fish measured by long-range, mid-frequency sonar,'' \emph{The
		Journal of the Acoustical Society of America}, vol. 141, no.~6, pp.
	4354--4371, 2017.
	
	\bibitem{abraham2002active}
	D.~A. Abraham and P.~K. Willett, ``Active sonar detection in shallow water
	using the page test,'' \emph{IEEE Journal of Oceanic Engineering}, vol.~27,
	no.~1, pp. 35--46, 2002.
	
	\bibitem{berg2018classification}
	H.~Berg and K.~T. Hjelmervik, ``Classification of anti-submarine warfare sonar
	targets using a deep neural network,'' in \emph{OCEANS 2018 MTS/IEEE
		Charleston}.\hskip 1em plus 0.5em minus 0.4em\relax IEEE, 2018, pp. 1--5.
	
	\bibitem{hjelmervik2017sonar}
	K.~T. Hjelmervik, D.~H.~S. Stender, H.~Berg \emph{et~al.}, ``Sonar scattering
	from the sea bottom near the {N}orwegian coast,'' in \emph{OCEANS
		2017-Anchorage}.\hskip 1em plus 0.5em minus 0.4em\relax IEEE, 2017, pp. 1--5.
	
	\bibitem{davey2007comparison}
	S.~J. Davey, M.~G. Rutten, and B.~Cheung, ``A comparison of detection
	performance for several track-before-detect algorithms,'' \emph{EURASIP
		Journal on Advances in Signal Processing}, vol. 2008, pp. 1--10, 2007.
	
	\bibitem{diamant2019active}
	R.~Diamant, D.~Kipnis, E.~Bigal, A.~Scheinin, D.~Tchernov, and A.~Pinchasi,
	``An active acoustic track-before-detect approach for finding underwater
	mobile targets,'' \emph{IEEE Journal of Selected Topics in Signal
		Processing}, vol.~13, no.~1, pp. 104--119, 2019.
	
	\bibitem{de2019automatic}
	G.~De~Magistris, P.~Stinco, J.~R. Bates, J.~M. Topple, G.~Canepa, G.~Ferri,
	A.~Tesei, and K.~Le~Page, ``Automatic object classification for low-frequency
	active sonar using convolutional neural networks,'' in \emph{OCEANS 2019
		MTS/IEEE SEATTLE}.\hskip 1em plus 0.5em minus 0.4em\relax IEEE, 2019, pp.
	1--6.
	
	\bibitem{robey1992cfar}
	F.~C. Robey, D.~R. Fuhrmann, E.~J. Kelly, and R.~Nitzberg, ``A {CFAR} adaptive
	matched filter detector,'' \emph{IEEE Transactions on aerospace and
		electronic systems}, vol.~28, no.~1, pp. 208--216, 1992.
	
	\bibitem{seo2019underwater}
	I.~Seo, S.~Kim, Y.~Ryu, J.~Park, and D.~S. Han, ``Underwater moving target
	classification using multilayer processing of active sonar system,''
	\emph{Applied Sciences}, vol.~9, no.~21, p. 4617, 2019.
	
	\bibitem{cowell2010separation}
	D.~M. Cowell and S.~Freear, ``Separation of overlapping linear frequency
	modulated ({LFM}) signals using the fractional fourier transform,''
	\emph{IEEE transactions on ultrasonics, ferroelectrics, and frequency
		control}, vol.~57, no.~10, pp. 2324--2333, 2010.
	
	\bibitem{jacob2009applications}
	R.~Jacob, T.~Thomas, and A.~Unnikrishnan, ``Applications of fractional fourier
	transform in sonar signal processing,'' \emph{IETE Journal of Research},
	vol.~55, no.~1, pp. 16--27, 2009.
	
	\bibitem{yu2017fractional}
	G.~Yu, S.-c. Piao, and X.~Han, ``Fractional fourier transform-based detection
	and delay time estimation of moving target in strong reverberation
	environment,'' \emph{IET Radar, Sonar \& Navigation}, vol.~11, no.~9, pp.
	1367--1372, 2017.
	
	\bibitem{stinco2021automatic}
	P.~Stinco, G.~De~Magistris, A.~Tesei, and K.~D. LePage, ``Automatic object
	classification with active sonar using unsupervised anomaly detection,'' in
	\emph{2020 28th European Signal Processing Conference (EUSIPCO)}.\hskip 1em
	plus 0.5em minus 0.4em\relax IEEE, 2021, pp. 46--50.
	
	\bibitem{barniv1985dynamic}
	Y.~Barniv, ``Dynamic programming solution for detecting dim moving targets,''
	\emph{IEEE Transactions on Aerospace and Electronic Systems}, vol. AES-21,
	no.~1, pp. 144--156, 1985.
	
	\bibitem{wei2018novel}
	M.~Wei, B.~Shi, C.~Hao, and S.~Yan, ``A novel weak target detection strategy
	for moving active sonar,'' in \emph{2018 OCEANS-MTS/IEEE Kobe Techno-Oceans
		(OTO)}.\hskip 1em plus 0.5em minus 0.4em\relax IEEE, 2018, pp. 1--6.
	
	\bibitem{jing2016detection}
	C.~Jing, Z.~Lin, and J.~Li, ``Detection and tracking of an underwater target
	using the combination of a particle filter and track-before-detect,'' in
	\emph{OCEANS 2016-Shanghai}.\hskip 1em plus 0.5em minus 0.4em\relax IEEE,
	2016, pp. 1--5.
	
	\bibitem{northardt2018track}
	T.~Northardt and S.~C. Nardone, ``Track-before-detect bearings-only
	localization performance in complex passive sonar scenarios: A case study,''
	\emph{IEEE Journal of Oceanic Engineering}, vol.~44, no.~2, pp. 482--491,
	2018.
	
	\bibitem{streit1994maximum}
	R.~L. Streit and T.~E. Luginbuhl, ``Maximum likelihood method for probabilistic
	multihypothesis tracking,'' in \emph{Signal and Data Processing of Small
		Targets 1994}, vol. 2235.\hskip 1em plus 0.5em minus 0.4em\relax
	International Society for Optics and Photonics, 1994, pp. 394--405.
	
	\bibitem{streit2002multitarget}
	R.~L. Streit, M.~L. Graham, and M.~J. Walsh, ``Multitarget tracking of
	distributed targets using histogram-{PMHT},'' \emph{Digital Signal
		Processing}, vol.~12, no. 2-3, pp. 394--404, 2002.
	
	\bibitem{vu2013track}
	H.~Vu, S.~Davey, F.~Fetcher, S.~Arulampalam, R.~Ellem, and C.~Lim,
	``Track-before-detect for an active towed array sonar,'' \emph{Acoustics.
		Victor Harbor, South Australia}, 2013.
	
	\bibitem{gaetjens2017histogram}
	H.~X. Gaetjens, S.~J. Davey, S.~Arulampalam, F.~K. Fletcher, and C.-C. Lim,
	``Histogram-{PMHT} for fluctuating target models,'' \emph{IET Radar, Sonar \&
		Navigation}, vol.~11, no.~8, pp. 1292--1301, 2017.
	
	\bibitem{jauffret1990track}
	C.~Jauffret and Y.~Bar-Shalom, ``Track formation with bearing and frequency
	measurements in clutter,'' in \emph{29th IEEE Conference on Decision and
		Control}.\hskip 1em plus 0.5em minus 0.4em\relax IEEE, 1990, pp. 3335--3336.
	
	\bibitem{willett2005application}
	P.~Willett and S.~Coraluppi, ``Application of the {MLPDA} to bistatic sonar,''
	in \emph{2005 IEEE Aerospace Conference}.\hskip 1em plus 0.5em minus
	0.4em\relax IEEE, 2005, pp. 2063--2073.
	
	\bibitem{blanding2007sequential}
	W.~Blanding, P.~Willett, and S.~Coraluppi, ``Sequential {ML} for multistatic
	sonar tracking,'' in \emph{OCEANS 2007-Europe}.\hskip 1em plus 0.5em minus
	0.4em\relax IEEE, 2007, pp. 1--6.
	
	\bibitem{schoenecker2013ml}
	S.~Schoenecker, P.~Willett, and Y.~Bar-Shalom, ``{ML--PDA} and {ML--PMHT}:
	Comparing multistatic sonar trackers for {VLO} targets using a new
	multitarget implementation,'' \emph{IEEE Journal of Oceanic Engineering},
	vol.~39, no.~2, pp. 303--317, 2013.
	
	\bibitem{korneliussen2003synthetic}
	R.~J. Korneliussen and E.~Ona, ``Synthetic echograms generated from the
	relative frequency response,'' \emph{ICES Journal of Marine Science},
	vol.~60, no.~3, pp. 636--640, 2003.
	
	\bibitem{stanton2010new}
	T.~K. Stanton, D.~Chu, J.~M. Jech, and J.~D. Irish, ``New broadband methods for
	resonance classification and high-resolution imagery of fish with
	swimbladders using a modified commercial broadband echosounder,'' \emph{ICES
		Journal of Marine Science}, vol.~67, no.~2, pp. 365--378, 2010.
	
	\bibitem{pailhas2010analysis}
	Y.~Pailhas, C.~Capus, K.~Brown, and P.~Moore, ``Analysis and classification of
	broadband echoes using bio-inspired dolphin pulses,'' \emph{The Journal of
		the Acoustical Society of America}, vol. 127, no.~6, pp. 3809--3820, 2010.
	
	\bibitem{love1971measurements}
	R.~H. Love, ``Measurements of fish target strength: a review,'' \emph{Fish.
		Bull}, vol.~69, no.~4, pp. 703--715, 1971.
	
	\bibitem{vinh2010information}
	N.~X. Vinh, J.~Epps, and J.~Bailey, ``Information theoretic measures for
	clusterings comparison: Variants, properties, normalization and correction
	for chance,'' \emph{The Journal of Machine Learning Research}, vol.~11, pp.
	2837--2854, 2010.
	
	\bibitem{misra2004spectral}
	H.~Misra, S.~Ikbal, H.~Bourlard, and H.~Hermansky, ``Spectral entropy based
	feature for robust {ASR},'' in \emph{2004 IEEE International Conference on
		Acoustics, Speech, and Signal Processing}, vol.~1.\hskip 1em plus 0.5em minus
	0.4em\relax IEEE, 2004, pp. I--193.
	
	\bibitem{ng2002spectral}
	A.~Y. Ng, M.~I. Jordan, and Y.~Weiss, ``On spectral clustering: Analysis and an
	algorithm,'' in \emph{Advances in neural information processing systems},
	2002, pp. 849--856.
	
\end{thebibliography}


\appendix

\section{Details of the Spectral Clustering Algorithm}\label{sec: Appendix}

For the sake of completeness, we now detail our implementation to the algorithm in \cite{ng2002spectral}.
\begin{enumerate}
	\item
	Define the matrix $\bf A$ according to \eqref{eq: A_ij};
	\item 
	Define the diagonal matrix  $\bf D$ according to \eqref{eq: D_i};
	\item 
	Calculate the symmetric graph-Laplacian by \eqref{eq: graph-Laplacian};
	\item
	Assuming the number of clusters is $K$, calculate the $K$ eigenvectors of $\bf L$ corresponding to the $K$ smallest eigenvalues (this is because we use \eqref{eq: graph-Laplacian} instead of ${\bf D}^{-1/2}{\bf A}{\bf D}^{-1/2}$ used in \cite{ng2002spectral}), stack them in the columns of a matrix ${\bf X}_{N \times K} $, and normalize its rows to get matrix $\bf Y$ with entries
	\begin{equation}
		Y_{ij} = \frac{X_{ij}}{ (\sum_{j=1}^N X_{ij}^2 )^{1/2}} \;;
	\end{equation}
	\item
	Refer to each row in $\bf Y$ as a point in $\mathbb{R}^K$, classify by K-means to one of $K$ clusters, and assign the cluster index associated with this row to the corresponding data-point.
\end{enumerate}

\end{document}